\documentclass[12pt,preprint]{aastex}

\def\HII{{\ion{H}{2}}}
\def\OII{[{\ion{O}{2}}]}
\def\4959_5007{[\ion{O}{3}]~$\lambda \lambda$4959,5007}
\def\OIII49595007{[\ion{O}{3}]~$\lambda \lambda 4959,5007$}
\def\ratioR23{([\ion{O}{2}]~$\lambda$3727 +
[\ion{O}{3}]~$\lambda\lambda$4959,5007)/H$\beta$}
\def\R23{${\rm R}_{23}$}
\def\dS23{${\rm S}_{23}$}
\def\OIIIl{[\ion{O}{3}]~$\lambda$5007}
\def\Zsun{${\rm Z}_{\odot}$}
\def\Msun{${\rm M}_{\odot}$}

\def\NII{[{\ion{N}{2}}]}
\def\OIIIOII{[\ion{O}{3}]/[\ion{O}{2}]}
\def\lOII{[\ion{O}{2}]~$\lambda \lambda$3726,9}
\def\lNII{[{\ion{N}{2}}]~$\lambda$6584}
\def\SIIl{[\ion{S}{2}]~$\lambda \lambda$6717,31}
\def\NIIOII{[\ion{N}{2}]/[\ion{O}{2}]}
\def\ratioNIIOII{[\ion{N}{2}]~$\lambda$6584/[\ion{O}{2}]~$\lambda$3727}
\def\OH{$\log({\rm O/H})+12$}
\def\NIISII{[\ion{N}{2}]/[\ion{S}{2}]}
\def\ratioNIISII{[\ion{N}{2}]~$\lambda$6584/[\ion{S}{2}]~$\lambda
\lambda$6717,31}
\def\SIIISII{[\ion{S}{3}]/[\ion{S}{2}]}
\def\ratioS23{([\ion{S}{2}]~$\lambda \lambda$6717,31 +
[\ion{S}{3}]~$\lambda\lambda$9069,9532)/H$\beta$}
\def\NIIHa{[\ion{N}{2}]/H$\alpha$}
\def\NIIOIII{[\ion{N}{2}]/[\ion{O}{3}]}
\def\SII{[{\ion{S}{2}}]}
\def\SIII{[{\ion{S}{3}}]}
\def\Hb{{H$\beta$}}
\def\O4363{[{\ion{O}{3}}]~$\lambda$4363}
\def\OIII{[{\ion{O}{3}}]}
\def\Ha{{H$\alpha$}}

\begin{document}

\title{Using Strong Lines to Estimate Abundances in
Extragalactic \HII\ Regions and Starburst Galaxies}
\author{L.J. Kewley, M.A. Dopita}

\affil{Research School of Astronomy and Astrophysics, Australian National
University} \authoraddr{Private Bag, Weston Creek PO, ACT 2611, Australia}

\begin{abstract}

We have used a combination of stellar population synthesis and
photoionization models to develop a set of ionization parameter
and abundance diagnostics based only on the use of the strong
optical emission lines.  These models are applicable to both
extragalactic \HII\ regions and star-forming galaxies.
We show that, because our techniques solve explicitly for both the
ionization parameter and the chemical abundance, the diagnostics
presented here are an improvement on earlier techniques based on
strong emission-line ratios. 

Our techniques are applicable
at all metallicities.  In particular, for metallicities above
half solar, the ratio \NIIOII\ provides a very reliable diagnostic
since it is ionization parameter independant and does not have a
local maximum.  This ratio has not been used historically because of
worries about reddening corrections. However, we show that the
use of classical reddening curves is quite sufficient to allow
this \NIIOII\ diagnostic to be used with confidence as a reliable
abundance indicator.

As we had previously shown, the commonly-used  abundance diagnostic
\R23\ depends strongly on the ionization parameter, while the
commonly-used ionization parameter diagnostic \OIIIOII\ depends
strongly on abundance. The iterative method of solution presented here
allows both of these parameters to be obtained without recourse to
the use of temperature-sensitive line ratios involving faint emission
lines.

We compare three commonly-used abundance diagnostic techniques and
show that individually, they contain systematic and random errors.
This is a problem affecting many abundance diagnostics and the errors
generally have not been properly studied or understood due to the lack
of a reliable comparison abundance, except for very low metallicities,
where the \OIII~$\lambda$4363 auroral line is used.  Here, we show
that the average of these techniques provides a fairly reliable
comparison abundance indicator against which to test new diagnostic
methods.

The cause of the systematic effects are discussed, and we present a
new `optimal' abundance diagnostic method based on the use of line ratios
involving \NII, \OII, \OIII, \SII\ and the Balmer lines. This combined
diagnostic appears to suffer no apparent systematic errors, can be used
over the entire abundance range and significantly reduces the
random error inherent in previous techniques.

Finally, we give a recommended procedure for the derivation of abundances
in the case that only spectra of limited wavelength coverage are
available so that the optimal method can no longer be used.

\end{abstract}

\keywords{interstellar: HII, abundances---galaxies: chemical evolution,
abundances, starburst}

\maketitle

\section{Introduction}

Powerful constraints on theories of galactic chemical evolution and on the
star formation histories of galaxies can be derived from the accurate
determination of chemical abundances either in individual star forming regions
distributed across galaxies or through the comparison of abundances
between galaxies.

The use of optical emission to estimate abundances in extragalactic
ionized hydrogen (\HII) regions dates back as early as the 1940s
\citep{Aller42}. The advent of linear detectors with high dynamic range,
and the development of quantitative photoionization models provided the
means whereby chemical abundances could be measured quantitatively in
both galactic and extragalactic \HII\ regions
\citep[eg.,][]{Peimbert75,Pagel86,Osterbrock89,Shields90,Aller90}.
In such work, measurements of the Hydrogen and Helium recombination lines
are used along with collisionally excited lines observed in one or more
ionization states of heavy elements.  Oxygen is commonly used
as the reference element because it is relatively abundant, it emits strong
lines in the optical regime, it is observed in several ionization states,
and line ratios of frequently-observed lines can provide good temperature
and density diagnostics (ie. \ion{O}{1}~$\lambda$6300,
\OII~$\lambda$3727,7318,7324 \OIII~$\lambda$4363,4959,5007). However,
the densities of extragalactic \HII\ regions are so low that
the density-sensitive line ratios are rarely used.  

Ideally, the oxygen abundance is measured directly from the ionic 
abundances obtained using a
determination of the electron temperature of the \HII\ region, and including
an appropriate correction factor to allow for the unseen stages of
ionization. This is known as the Ionization Correction Factor (ICF)
method.  Electron temperature is useful as an abundance indicator
since higher chemical abundances increase nebular cooling, leading to lower
\HII\ region temperatures.  The electron temperature can be determined
from the ratio of the auroral line \O4363\ to a lower excitation line
such as \OIIIl. In practice, however, \O4363\ is often very weak, even
in metal-poor environments, and cannot be observed at all in higher
metallicity galaxies. In addition, \O4363\ may be subject to systematic
errors when using global spectra.  For example, \citet{Kobulnicky99b}
found that for low metallicity galaxies, the \O4363\ diagnostic
systematically underestimated the global oxygen abundance, while for
more massive, metal-rich galaxies, empirical calibrations using strong
emission-line ratios can serve as reliable indicators of the overall
oxygen abundance in \HII\ regions.

With the rapidly increasing interest in the star formation history of the
high-redshift universe, it has become much more important to determine the
chemical evolutionary state of distant galaxies, even when only  spatially
unresolved (global) emission-line spectra are available
\citep{Steidel96, Kobulnicky99a, Kobulnicky99b}. However given the range of
temperatures, ionization parameters, and metallicities encountered in different
galactic systems, abundances derived using global spectra may often
be subject to important systematic errors.

For such galaxies and other star-forming regions without a measurable
[{\ion{O}{3}}] $\lambda$4363 flux, abundance determinations become entirely
dependent on the measurement of the ratios of the stronger emission-lines.
These still have the potential to deliver reasonably accurate estimates.
The most commonly used such ratio is \ratioR23 (otherwise known as \R23),
first proposed by \citet{Pagel79}. The logic for the use of this ratio is
that it provides an estimate of the total cooling due to oxygen, which,
given that oxygen is one of the principle nebular coolants, should in turn be
sensitive to the oxygen abundance. However, the calibration of this ratio
in terms of the abundance has proved to be rather
difficult due to the lack of good-quality independent data.  As discussed
above, \O4363\ is usually well-observed only for metal-poor (Z$<$0.5 \Zsun)
star-forming regions.  At higher metallicities, detailed theoretical
model fits to the data must be used.  As a result, many calibrations
of \R23 are available, including \citet{Pagel79,Pagel80,Edmunds84,McCall85,
Dopita86,Torres-Peimbert89,Skillman89,McGaugh91,Zaritsky94,Pilyugin00},
Charlot01.
Recent calibrations of \R23 produce oxygen abundances which are
comparable in accuracy to direct methods relying on the measurement of
nebular temperature, at least in the cases where these direct methods are
available for comparison \citep{McGaugh91}.
Rather than calculating model fits to the data, some authors have
favoured a more empirical approach.  For example,
\citet{Thurston96} estimated the electron temperature from
the easier-to-observe \NII\ line, which was then calibrated against
\R23. However, this has the drawback that the strength of the \NII\ must
be calibrated separately, and we are dealing here with an element which
has both primary and secondary nucleosynthesis components whose relative
contributions depend on metallicity.

One drawback of using \R23 (and many of the other emission-line abundance
diagnostics) is that it depends also on the ionization parameter ($q$) defined
here as:
\begin{equation}
q=\frac{S_{{\rm H}^{0}}}{n}  \label{1}
\end{equation}
where $S_{{\rm H}^{0}}$ is the ionizing photon flux through a unit 
area, and $n$
is the local number density of hydrogen atoms. The ionization parameter $q$, can be physically
interpreted as the maximum velocity of an ionization front that can be
driven by the local radiation field. This local ionization parameter can be
made dimensionless by dividing by the speed of light to give the more
commonly used ionization parameter; ${\cal U}\equiv q/c.$ Some
calibrations have attempted to take this into account
\citep[eg.][]{McGaugh91}, but others do not \citep[eg.][]{Zaritsky94}.
Another difficulty in the use of \R23 and many other emission-line
abundance diagnostics is that they are double valued in terms of the abundance.
This is because at low abundance the intensity of the forbidden
lines scales roughly with the chemical abundance while at high abundance the
nebular cooling is dominated by the infrared fine structure lines and
the electron temperature becomes too low to collisionally excite the optical
forbidden lines. When only double-valued diagnostics are available,
an iterative approach which explicitly solves for the ionization
parameter also helps to resolve the abundance ambiguities, as will be
described in this paper.

A combination of an extended IUE database, IRAS data, and the existence of
better stellar evolutionary tracks which include mass loss and overshooting
allowed large advances in the modelling of starburst emission spectra
in recent years \citep{Mas-Hesse91}.  With recent advances in nebular
physics, detailed photoionization models have been produced
which include self-consistent treatment of nebular and
dust physics \citep[eg.,][]{Dopita96,Ferland98}.  These can be used in
conjunction with modern stellar population synthesis models to synthesize the
spectra of starburst galaxies from the UV to the X-ray regime.  These models
have been used by us \citep{Dopita00a,Kewley01} to simulate the
emission-line spectra of \HII\ regions and starburst galaxies, respectively.
Here, these models are used to develop an optimal scheme for abundance
determination based on the range of possible combinations of bright optical
or IR emission-lines which are likely to be available to the observer.

\section{Models}
\label{models}

The models used here are those described in \citet{Dopita00a}.  Briefly,
the stellar population synthesis codes PEGASE \citep{Fioc97}
and STARBURST99 \citep{Leitherer99} were used to generate the ionizing EUV radiation
field.  We use instantaneous burst models at zero-age with a Saltpeter IMF
and lower and upper mass limits of 0.1 and 120~\Msun\ respectively.
Metallicities were varied from 0.05 to 3~$\times$~solar. This range is
restricted by the metallicities available in the population synthesis models.
In \citet{Dopita00a}, we show that the ionizing radiation fields are
identical for the two codes, and that the zero-age instantaneous burst models
are the most appropriate to use to model HII regions, when compared with
continuous burst models.

The ionizing EUV radiation fields were input into our photoionization and
shock code, MAPPINGS~III \citep{Sutherland93}.  We used plane parallel,
isobaric models with $P/k$ = 10$^{5}$ cm$^{-3}$K (for electron temperatures
of 10$^{4}$K. This pressure corresponds to a density of order 10~cm$^{-3}$;
typical of giant extragalactic \ion{H}{2} regions).  Dust physics is treated
explicitly through absorption, grain charging and photoelectric heating and
the use of a standard MRN \citep{Mathis77} grain size distribution for both
carbonaceous and silicaceous grain types. We took the undepleted solar
abundances to be those of \citet{Anders89}. These abundances and the
gas-phase depletion factors adopted for each element are shown in
Table~\ref{Z_table}. For non-solar metallicities we assume that both the
dust model and the depletion factors are unchanged, since we have no way
of estimating what they may be otherwise.  All elements except nitrogen
and helium are taken to be primary nucleosynthesis elements.  For helium,
we assume a primary nucleosynthesis component in addition to the primordial
value derived from \cite{Russell92}.
This primary component is matched empirically to provide the observed
abundances at SMC, LMC and solar abundances; \cite{Anders89}, \cite{Russell92}.
Nitrogen is assumed to be a secondary nucleosynthesis element above
metallicities of $0.23$ solar, but as a primary nucleosynthesis element at
lower metallicities. This is an empirical fit to the observed behaviour of
the N/O ratio in \ion{H}{2} regions \citep{VanZee97}.

Photoionization models were calculated for the range of metallicities
from 0.05 to 3$\times$\Zsun and for ionization parameter (as defined in the
Introduction) ranging from $q=5\times 10^{6}\,{\rm cm\,s}^{-1}$
to $q=3\times 10^{8}\,{\rm cm\,s}^{-1}$ ($\log({\cal U}) = -3.78$ to $-2$).
For our metallicities of 0.05, 0.1, 0.2, 0.5, 1.0, 1.5, 2.0, 
3.0 times solar, the corresponding metallicities in log(O/H)+12 are 
7.6, 7.9, 8.2, 8.6, 8.9, 9.1, 9.2, 9.4.  Note that for the new value of
solar abundance given in \citet{Allende01},
our metallicities become 0.09, 0.2, 0.4, 0.9, 1.7, 2.6, 3.5, 5.2 times 
solar.

\section{Diagnostic Diagrams for the Ionization Parameter}
\label{ionization}

Assuming that the metallicity and the shape of the EUV spectrum are defined,
the local ionization state in an \HII\ region is characterized by the
local ionization parameter. All models with a similar ionization parameter
will then produce very similar spectra. For plane-parallel
photoionization models, this ionization parameter defined at the inner
surface of the nebular ionized slab defines a unique ionization structure,
and therefore a unique emission line spectrum for a given metallicity and
EUV radiation field. In a spherical nebula, models are not unique because
of the spherical divergence of the radiation field. What is usually done
is to define an effective mean ionization parameter in terms of the
Str\"{o}mgren radius, $R_{{\rm s}}$:
\begin{equation}
q_{{\rm eff}}=\frac{Q_{{\rm H}^{0}}}{4\pi R_{{\rm s}}^{2}n}  \label{2}
\end{equation}
where $Q_{{\rm H}^{0}}$ is the flux of ionizing photons produced by the
exciting stars above the Lyman limit. If the thermal gas only occupies a
fraction of the available volume, this definition must be modified
accordingly.

\begin{figure}
\plotone{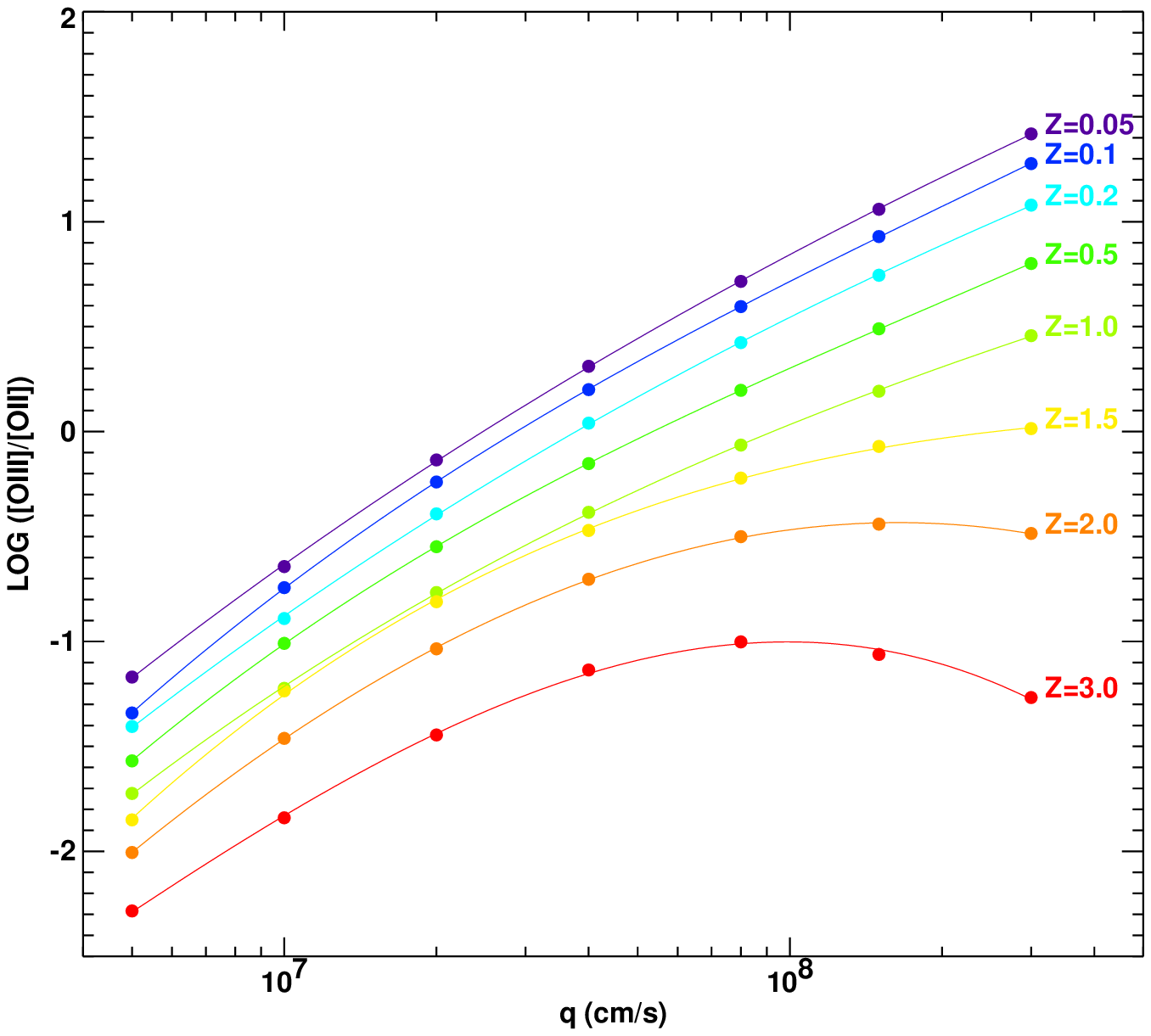}
\caption[f1.eps]{The 
log([\ion{O}{3}]~$\lambda 5007/$[\ion{O}{2}]~$\lambda \lambda 3726,9$) diagnostic  versus
ionization
parameter.  Curves
for each metallicity between $Z = 0.05$ to $3.0$ \Zsun\
are shown.  Filled circles represent the data points from our
models at ionization parameters from left to right of $5\times 10^{6}$,
$1\times 10^{7}$, $2\times 10^{7}$, $4\times 10^{7}$, $8\times 10^{7}$,
$1.5\times 10^{8}$, $3\times 10^{8}$ cm/s.  This figure is available in
color from the on-line version of this article.
\label{OIIIOII_vs_q}}
\end{figure}

Many of the abundance diagnostics commonly used are sensitive to
the ionization parameter, and for some ranges of metallicity, they
are not useful unless the ionization parameter can be constrained
within a small range of possible values.  Provided that the EUV spectrum
of the exciting source is reasonably well constrained, the ionization
parameter is  best determined using the ratios of emission-lines of
different ionization stages of the same element.  In general, the larger
the difference in ionization potentials of the two stages, the better
the ratio will constrain the ionization parameter. A commonly used
ionization parameter diagnostic is based on the ratio of the \OIIIl\ and
\lOII\ emission line fluxes; Figure~\ref{OIIIOII_vs_q}.

\placefigure{OIIIOII_vs_q}

The smooth curves are third order polynomial fits to the models for each
set of metallicities.  They have the form:

\begin{equation}
R=k_{0} + k_{1}x + k_{2}x^{2} + k_{3}x^{3}  \label{eq_OIIIOII}
\end{equation}

where $R$ is the flux ratio, in this case, log(\OIIIOII), $k_{0-3}$ are
constants given in Table~\ref{qcoefftable}, and $x$ is the ionization
parameter of the form $\log (q)$.

\placetable{qcoefftable}

It is evident that \OIIIOII\ is not only sensitive to the ionization
parameter, but is also strongly dependant on metallicity. An
initial guess of metallicity is essential in order to allow a suitable
constraint on the ionization parameter. This estimate could then be
used in order to then obtain a second, more accurate estimate of the
metallicity, and this iterative procedure can be continued
until convergence on both parameters is obtained.

\begin{figure}
\plotone{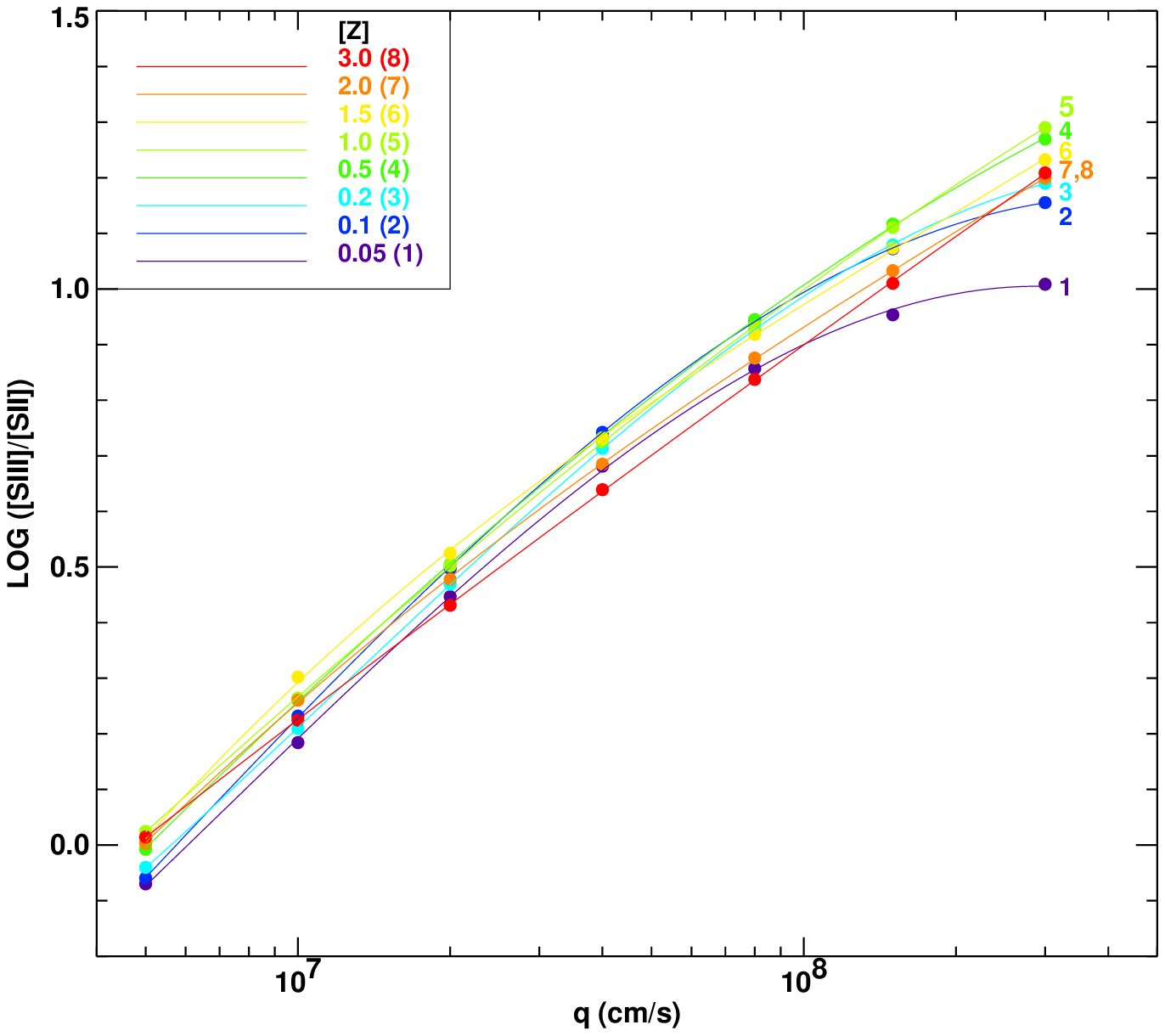}
\caption[f2.eps]{The log([\ion{S}{3}]~$\lambda \lambda 9069,9532$/
[\ion{S}{2}]~$\lambda \lambda 6717,31$)
 diagnostic versus ionization parameter.  Curves
for each metallicity between $Z = 0.05$ to $3.0$ \Zsun
are shown.  Filled circles represent the data points from our
models at ionization parameters from left to right of $5\times 10^{6}$,
$1\times 10^{7}$, $2\times 10^{7}$, $4\times 10^{7}$, $8\times 10^{7}$,
$1.5\times 10^{8}$, $3\times 10^{8}$ cm/s.  This figure is available in
color from the on-line version of this article.
\label{SIIISII_vs_q}}
\end{figure}

If, however, instead of the oxygen line ratios, one has reliable fluxes
for [\ion{S}{3}]~$\lambda$9069 and/or [\ion{S}{3}]~$\lambda$9532 as well
as the \SIIl\ emission lines, then the [\ion{S}{3}]/[\ion{S}{2}]\ ratio
provides a rather more useful ionization parameter diagnostic, as shown
in Figure~\ref{SIIISII_vs_q}.

\placefigure{SIIISII_vs_q}

The lack of a strong metallicity dependence in this line ratio
is largely due to the fact that the lines used are both in the red
portion of the spectrum, and therefore remain strong to much higher
metallicities than the \lOII\ lines. The \lOII\ lines can only be excited by
relatively hot electrons, and so disappear in cooler, high-metallicity
\ion{H}{2} regions. The lack of a metallicity dependence indicates that the
[\ion{S}{3}]/[\ion{S}{2}] line ratio should make a rather good 
diagnostic of the
ionization parameter.  An iterative solution only becomes necessary at
the highest values of the [\ion{S}{3}]/[\ion{S}{2}] ratio; above
$\log$([\ion{S}{3}]/[\ion{S}{2}])$> 1.0$.

\section{Abundance Sensitive Diagnostic Diagrams}

In this section, we present a range of diagnostic diagrams for
determining abundance.  The choice of the appropriate diagram or
combination of diagrams that are to be used by the observer
depends on which of the bright emission lines are available.
In some cases, an initial guess of metallicity
is also useful.

\subsection{The \NIIOII\ Diagnostic Diagram}

\begin{figure}
\plotone{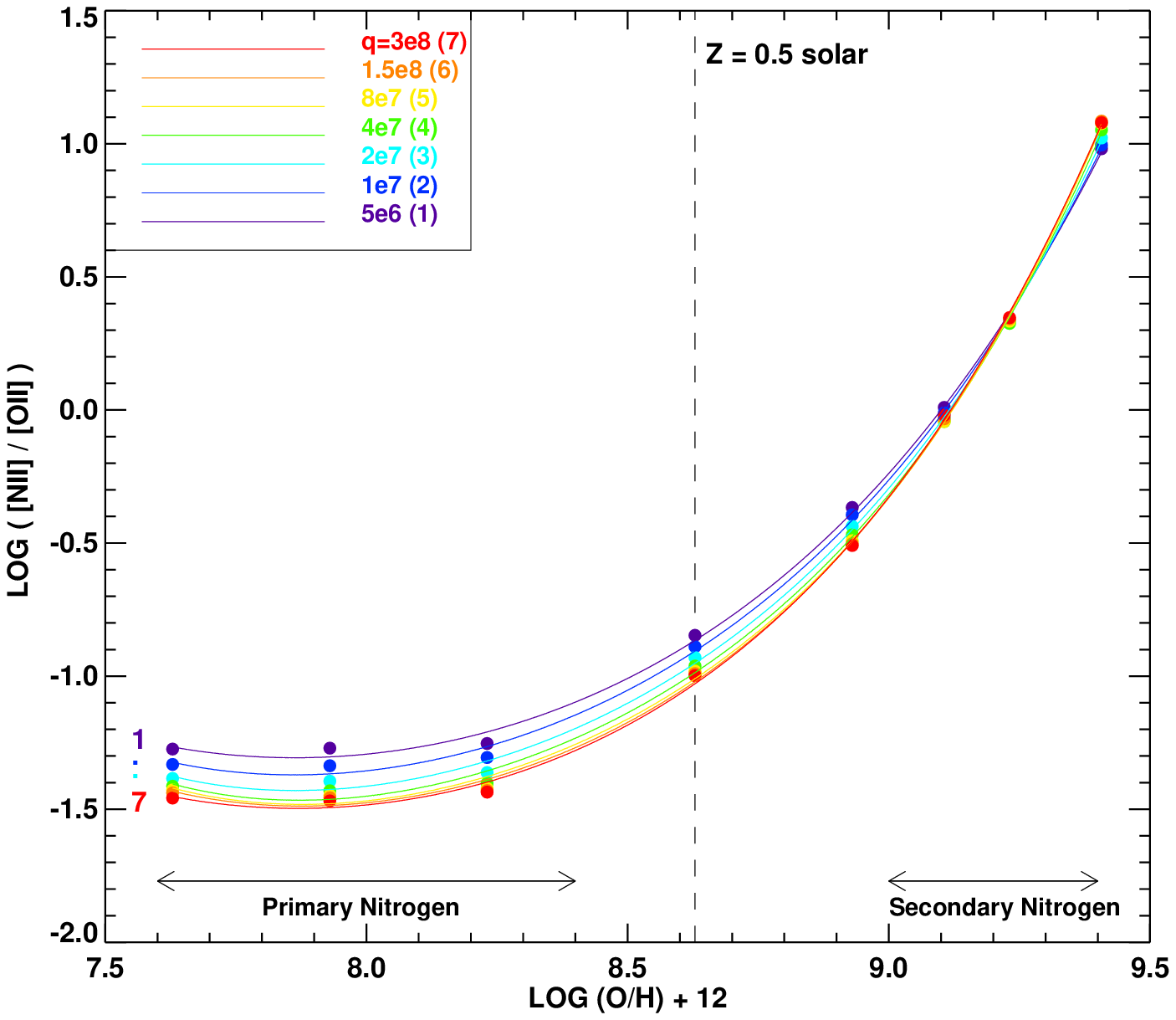}
\caption[f3.eps]{The log([\ion{N}{2}]~$\lambda 6584$/[\ion{O}{2}]~$\lambda \lambda 3726,9$) diagnostic for abundance versus
metallicity.  Curves
for each ionization parameter between ${\rm q} = 5\times10^{6}$ to
$3\times10^{8}$ cm/s
are shown.  Filled circles represent the data points from our
models at metallicities from left to right of 0.05, 0.1, 0.2,
0.5, 1.0, 1.5, 2.0, 3.0 \Zsun.
This figure is available in
color from the on-line version of this article.
\label{NIIOII_vs_Z}}
\end{figure}

\placefigure{NIIOII_vs_Z}

Our calibration of the \ratioNIIOII\ ratio as a function of
oxygen abundance
is shown in Figure~\ref{NIIOII_vs_Z} for a set of ionization parameters
varying from ${\rm q} = 5\times10^{6}$ to $3\times10^{8}$ cm/s .
The advantages of using the \NII\ and [\ion{O}{2}] lines are that they are
not affected
by underlying stellar populations and that they are quite strong,
even in low S/N spectra.  They do however, need to be corrected
for dust extinction.

The curves shown in Figure~\ref{NIIOII_vs_Z}
are fits to the models based on fourth order polynomials of the form:

\begin{equation}
\label{poly4}
\log(R)=k_{0} + k_{1}x + k_{2}x^{2} + k_{3}x^{3} + k_{4}x^{4}
\end{equation}

\noindent
where $R$ is the flux ratio, in this case, \NIIOII, $k_{0-4}$ are
constants given in Table~\ref{coefftable}, and the variable $x$ is
the metallicity given in the form \OH.

Figure~\ref{NIIOII_vs_Z} clearly shows that for Z $>$ 0.5 \Zsun$\,\,$
   (\OH\ $\gtrsim 8.6$), \NIIOII\ is a very useful diagnostic.
Because ${\rm N}^{+}$ and ${\rm O}^{+}$ have similar ionization
potentials, this ratio is almost independant of ionization parameter.
However, it increases very strongly with metallicity for two reasons.
Firstly, \NII\ is predominantly a secondary element
above metallicities of Z $>$ 0.5 \citep{Alloin79,Considere00} and
consequently the \NII\ flux scales more strongly than the \OII\ line
with increasing metallicity until about 2.0 - 3.0 \Zsun, when the
high metallicity and consequent low electron temperature in the nebula
makes both lines very difficult to observe.
Secondly, at high metallicity, the lower electron temperature gives
fewer thermal electrons of high energy, leading to a strong decrease
in the number of collisional excitations of the blue \OII\ lines
(which has a relatively high threshold energy for excitation) relative
to the lower-energy \NII\ lines.

For Z $<$ 0.5 \Zsun\ (\OH $< 8.6$), the metallicity dependence
of the \NIIOII\ ratio is lost because nitrogen (like oxygen) is predominantly
a primary nucleosynthesis element in this metallicity range.  In addition,
the nitrogen to oxygen abundance ratio shows large scatter from 
object to object.  In this regime, nitrogen 
production increases as a function of time since the bulk of the star
formation occurred \citep{Edmunds78,Matteucci85, Dopita97}.  Therefore, for a sample of \HII\ regions,
the varying age distribution of the stellar population from object to 
object will cause scatter in the N/O ratios observed.  
Good evidence for the primary dependence of nitrogen at low abundances
can be found in  Figure~4 of \citet{Considere00}, in
which $\log ({\rm N/O})$ derived from a large sample of \HII\ regions 
is compared
with the expected relations for a primary, secondary, and primary + secondary
origin for nitrogen production \citep{Vila-Costas93}.

We conclude that \NIIOII\ provides an excellent abundance diagnostic for
Z $>$ 0.5 \Zsun, but this ratio cannot be used at lower abundances.

For Z $>$ 0.5 \Zsun, the curves in Figure~\ref{NIIOII_vs_Z} can be
fit by a simple quadratic, facilitating abundance determination using
the \NIIOII\ ratio, ie

\begin{equation}
\log({\rm O/H}) + 12 =  \log[1.54020+1.26602\,{\rm R}
                             +0.167977\,{\rm R}^2] + 8.93
\end{equation}

where R is $\log($\NIIOII), and $\log({\rm O/H}) +12$ must be $\ge 8.6$
for this formula to give a reliable abundance.

\placetable{coefftable}

Given the large separation in wavelength between the \lOII\ and \lNII\
lines, the size and accuracy of determination of the reddening correction
remains a concern in the use of this diagnostic. However, we will show
in Section 6 that reddening correction using the Balmer decrement and
the classical reddening curves provides an adequate accuracy for the
use of the \NIIOII\ ratio in abundance determination. If there was any
grounds for concern about the reddening correction used, then a direct
measurement of the  \lOII\ to H{$\delta$} and \lNII\ to H{$\alpha$} ratios
followed by a correction using the theoretical Case B H{$\delta$} to
H{$\alpha$} ratio ($\sim 11$) would provide an adequate reddening correction.
An uncorrected reddening of E(B-V) of 0.5 will cause the abundance to be
over-estimated by roughly 0.5 dex in $\log({\rm O/H})+12$.

\subsection{The \NIISII\ Diagnostic Diagram}

\begin{figure}
\plotone{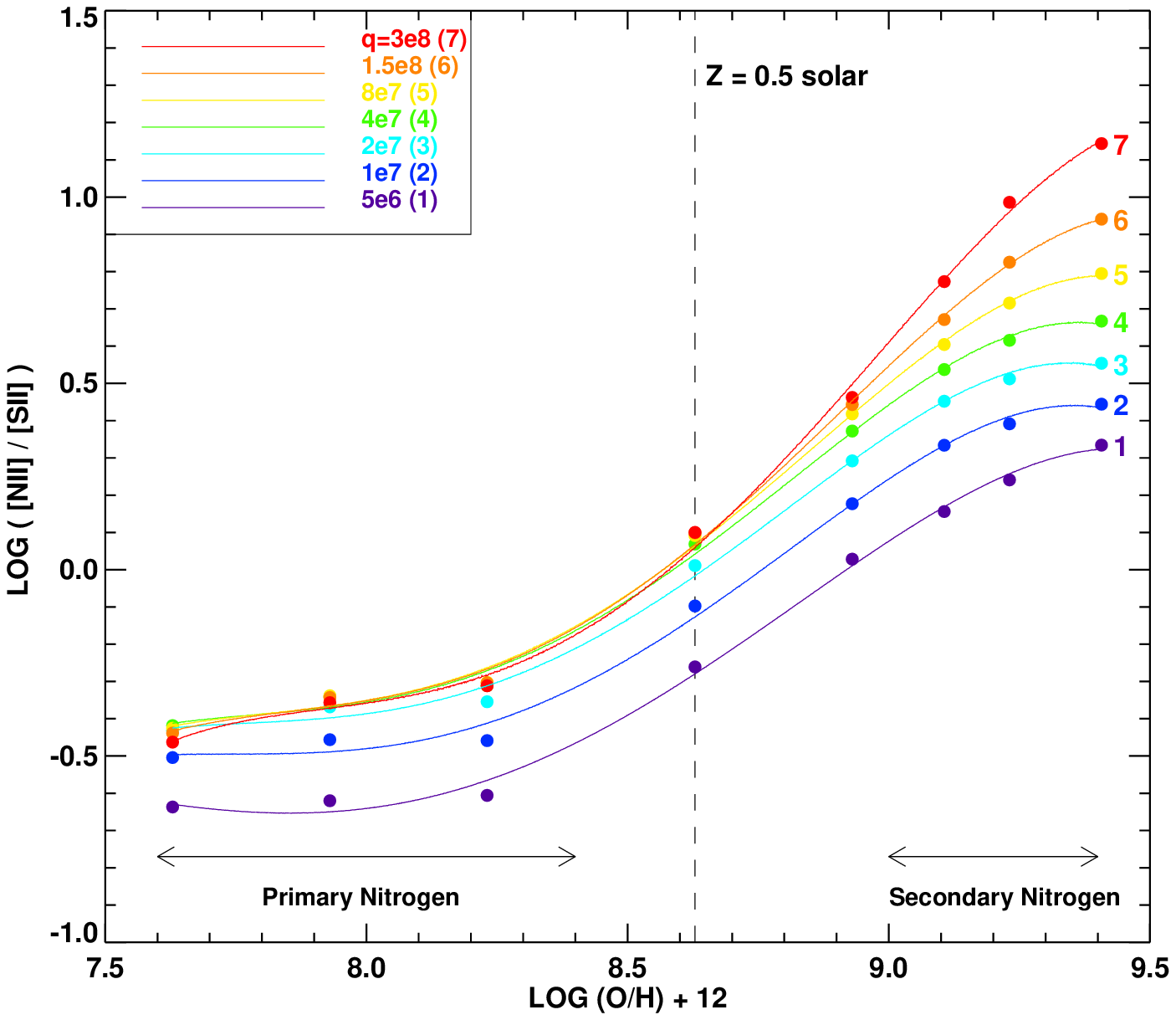}
\caption[f4.eps]{The log([\ion{N}{2}]~$\lambda 6584$/[\ion{S}{2}]~$\lambda \lambda 6717,31$) diagnostic for abundance versus
metallicity.  Curves
for each ionization parameter between ${\rm q} = 5\times10^{6}$ to
$3\times10^{8}$ cm/s
are shown.  Filled circles represent the data points from our
models at metallicities from left to right of 0.05, 0.1, 0.2,
0.5, 1.0, 1.5, 2.0, 3.0 \Zsun. This figure is available in
color from the on-line version of this article.
\label{NIISII_vs_Z}}
\end{figure}

Our calibration of \ratioNIISII\
is shown in Figure~\ref{NIISII_vs_Z}.

As in other diagrams, the smooth curves are fourth order polynomial
fits defined as in equation~\ref{poly4}, and the coefficients are
given in Table~\ref{coefftable}.

At high metallicity, nitrogen is a secondary nucleosynthesis element
and sulphur is a primary $\alpha$-process element.  Both lines are close
to each other in wavelength, and therefore have similar excitation
potentials.  At high metallicity, therefore, this line ratio is a function
of metallicity thanks primarily to the different nucleogenic status of the
two elements.  At low metallicity, both the elements are primary and the ratio
becomes insensitive to metallicity. This diagnostic is not as useful as
\NIIOII\ for the determination of abundance, but it has the advantage of
being far less sensitive to reddening.  As \NIISII\ is also dependant
on ionization parameter, one of the ionization parameter diagnostics
should be used in combination with this diagnostic.

\placefigure{NIISII_vs_Z}

\subsection{The \R23 Diagnostic Diagram}

\begin{figure}
\plotone{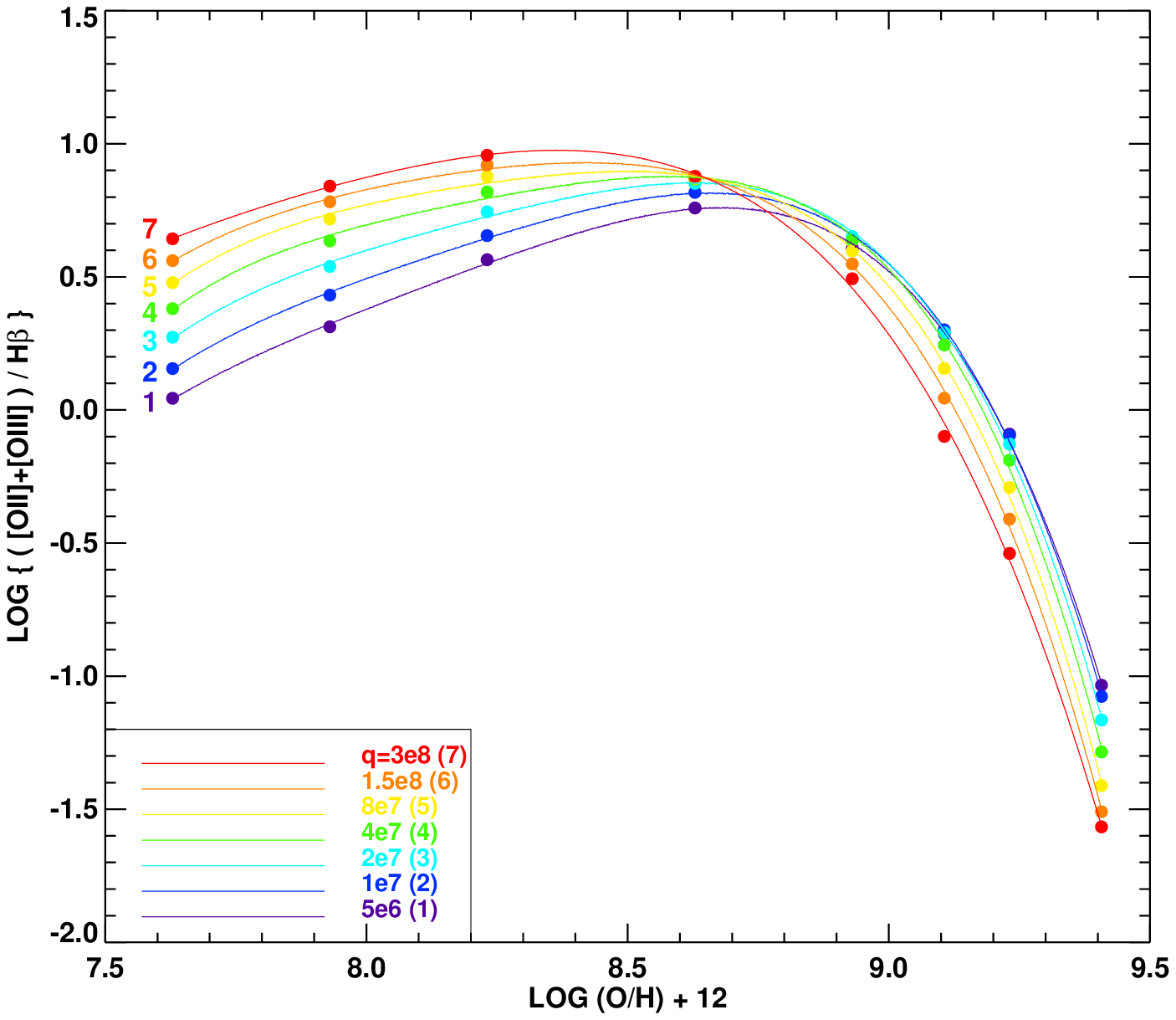}
\caption[f5.eps]{The $\log$(\ratioR23) (\R23) diagnostic for abundance
versus metallicity. Curves
for each ionization parameter between ${\rm q} = 5\times10^{6}$ to
$3\times10^{8}$ cm/s are shown.  Filled circles represent the data points
from our models at metallicities from left to right of 0.05, 0.1, 0.2,
0.5, 1.0, 1.5, 2.0, 3.0 \Zsun.  This figure is available in
color from the on-line version of this article.
\label{R23_vs_Z}}
\end{figure}

Since they were first introduced by \citet{Pagel79}, diagnostics based on the
so-called \R23 ratio, defined as \R23$ = $\ratioR23 have found extensive
application in the literature. Examples include
\citet{Alloin79,Edmunds84,McCall85,Dopita86,Pilyugin00,McGaugh91,Pilyugin01}
and \citet {Charlot01}. As explained in the introduction, \R23 is sensitive
to abundance, but is two-valued as a function of metallicity, 
reaching a maximum
at somewhat less than solar abundance (see Figure~\ref{R23_vs_Z}).
The smooth curves are fourth order polynomial fits to the models
for each ionization parameter defined as in equation~\ref{poly4}, with
coefficients listed in Table~\ref{coefftable} .

\placefigure{R23_vs_Z}

Because this ratio is two-valued with metallicity, the key problem
associated with the use of this diagnostic to derive abundance
is to determine which solution branch applies.
This is best determined by the use of an initial guess of the metallicity
based on an alternative diagnostic such as \NIIOII .

As has been shown in many previous studies, \R23 depends on the
ionization parameter, particularly for Z $< 0.5$ \Zsun,
but is less sensitive to metallicity in these ranges.

With such a diagnostic, a narrow range of possible ionization parameters can
be found using an ionization parameter sensitive ratio such as
\OIIIOII\, and then solving for metallicity.  Other methods have used
an empirical `correction factor' or `excitation parameter' to
correct the observed \R23 for ionization parameter, such as in
\citet{Pilyugin00,Charlot01}.  We prefer to estimate the ionization
parameter explicitly, based upon the theoretical models.

\subsection{The ${\rm S}_{23}$ Diagnostic Diagram}

\begin{figure}
\plotone{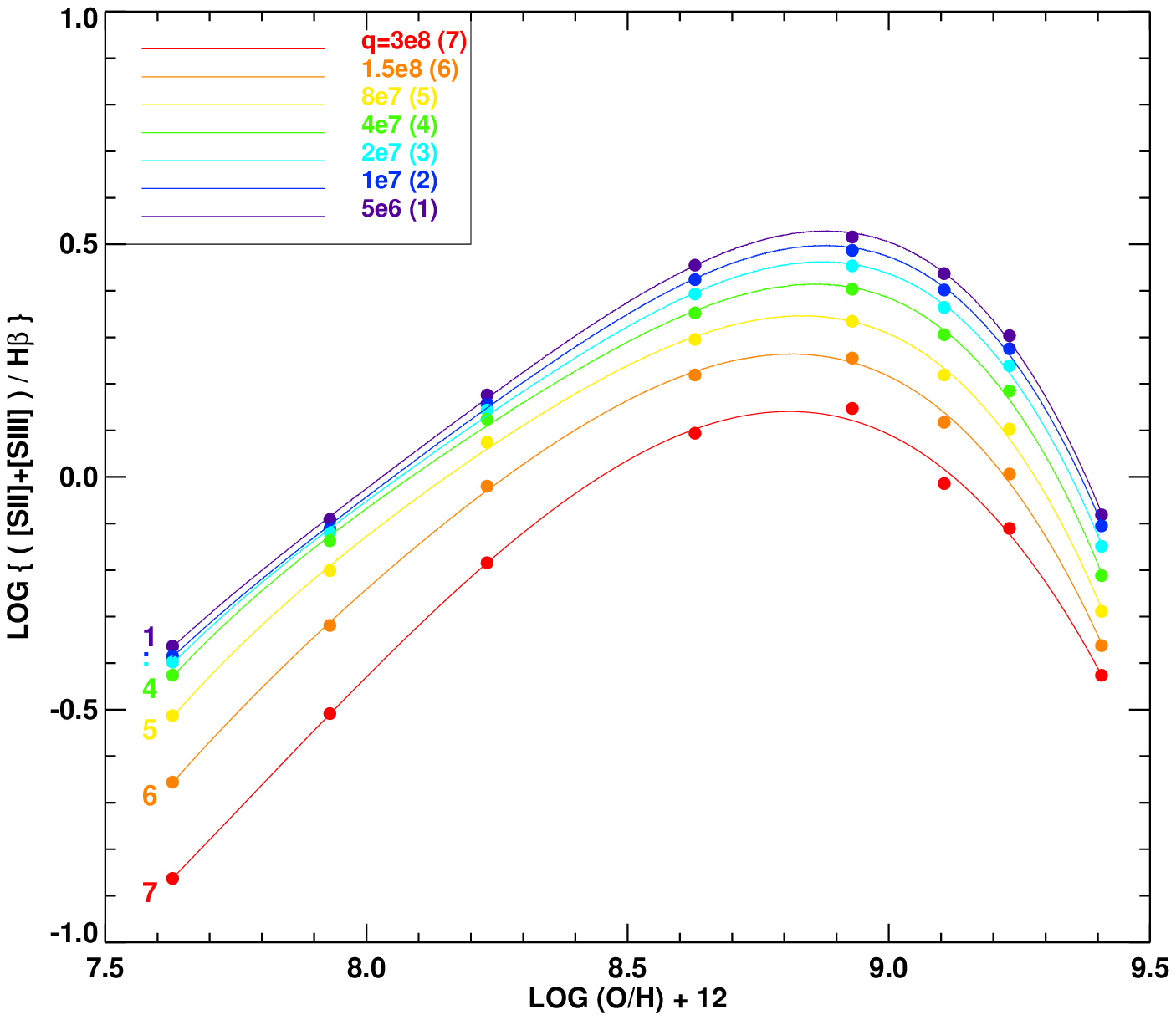}
\caption[f6.eps]{The $\log$(\ratioS23) (${\rm S}_{23}$) diagnostic for abundance
versus metallicity.  Curves
for each ionization parameter between ${\rm q} = 5\times10^{6}$ to
$3\times10^{8}$ cm/s
are shown.  Filled circles represent the data points from our
models at metallicities from left to right of 0.05, 0.1, 0.2,
0.5, 1.0, 1.5, 2.0, 3.0 \Zsun. This figure is available in
color from the on-line version of this article.
\label{S23_vs_Z}}
\end{figure}

Our calibration of the ratio \ratioS23
(popularly known as ${\rm S}_{23}$) is shown in Figure~\ref{S23_vs_Z}.
Again, we fit fourth order polynomials defined as in
equation~\ref{poly4}, and the coefficients are given in
Table~\ref{coefftable}.

\placefigure{S23_vs_Z}

As is the case in all such ratios of forbidden to recombination lines,
the ${\rm S}_{23}$ ratio has a maximum at a certain metallicity, and
therefore is two valued at all other metallicities.  For this particular
ratio the maximum occurs at a somewhat higher abundance than for
the \R23 ratio; at metallicities of roughly solar (\OH $\sim 8.8$).
Again, to raise the degeneracy in the solutions, an initial
guess of the metallicity must first be obtained from an alternative
diagnostic.

${\rm S}_{23}$ is quite dependant on ionization parameter for all
metallicities, and therefore the ionization parameter derived from the
   [\ion{S}{3}]/[\ion{S}{2}]$\,$ diagnostic should be used to eliminate
this as a free variable.

\subsection{The \NIIHa\ Diagnostic Diagram}

\begin{figure}
\plotone{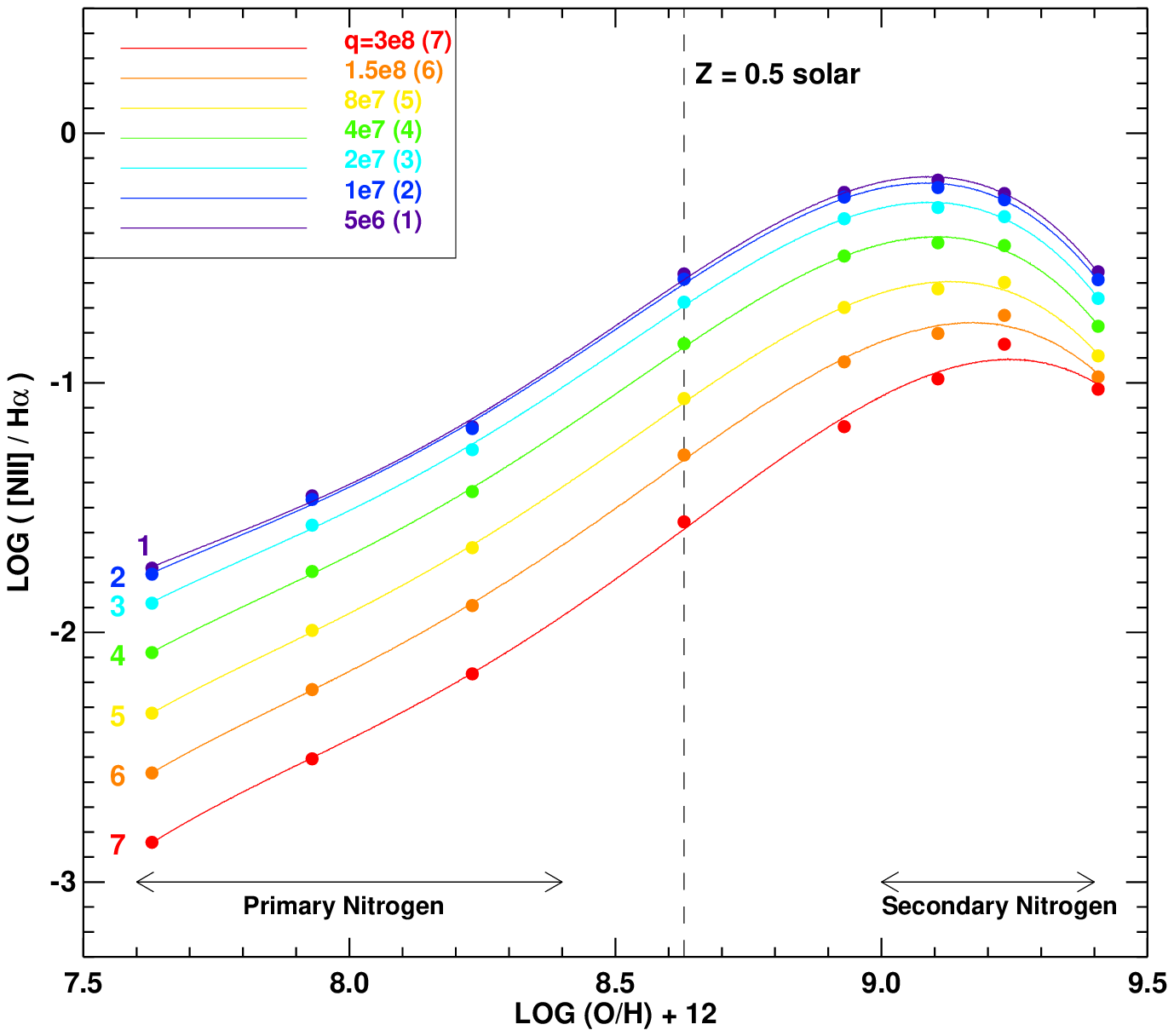}
\caption[f7.eps]{The log([\ion{N}{2}]~$\lambda 6584$/H$\alpha$) 
diagnostic 
for abundance versus
metallicity.  Curves
for each ionization parameter between ${\rm q} = 5\times10^{6}$ to
$3\times10^{8}$ cm/s
are shown.  Filled circles represent the data points from our
models at metallicities from left to right of 0.05, 0.1, 0.2,
0.5, 1.0, 1.5, 2.0, 3.0 \Zsun. This figure is available in
color from the on-line version of this article.
\label{NIIHa_vs_Z}}
\end{figure}

In the absence of other emission lines, the \NIIHa\ line ratio
can be used as a crude estimator of metallicity.
Our calibration of this ratio is shown in Figure~\ref{NIIHa_vs_Z},
with polynomial fits as in equation~\ref{poly4} and coefficients
given in Table~\ref{coefftable}

\placefigure{NIIHa_vs_Z}

At very low metallicity, this ratio scales simply as the nitrogen
abundance, to first order. However, it is known that in this
metallicity regime, the nitrogen abundance shows a lot of scatter
relative to the oxygen abundance, since the nitrogen abundance is much
more sensitive to the history of star formation in the galaxy considered.
As a result, this ratio is probably not very useful to estimate oxygen
abundance except as a means of determining the solution branch for the later
application of a ratio such as \R23. When the secondary production of
nitrogen dominates, at somewhat higher metallicity, the \NIIHa\ line ratio
continues to increase, despite the decreasing electron temperature.
Eventually, at still higher metallicities, nitrogen becomes the dominant
coolant in the nebula, and the electron temperature falls sufficiently
to ensure that that the nitrogen line weakens with increasing metallicity.
Since the [\ion{N}{2}] line is produced in the low-excition zone of
the \HII\ region, \NIIHa\ is also sensitive to ionization parameter.

\subsection{The \NIIOIII\ Diagnostic Diagram}

The advantage of using \NII\ and [\ion{O}{3}] lines is that they are
unaffected by absorption lines originating from underlying stellar
populations, they lie close to Balmer lines that can be used to eliminate
errors due to dust reddening, and they are both strong and easily observable
in the optical.  Both empirical and theoretical relationships for the \NIIOIII\
ratio as a function of oxygen abundance currently exist
\citep[eg.,][]{Considere00}.

\begin{figure}
\plotone{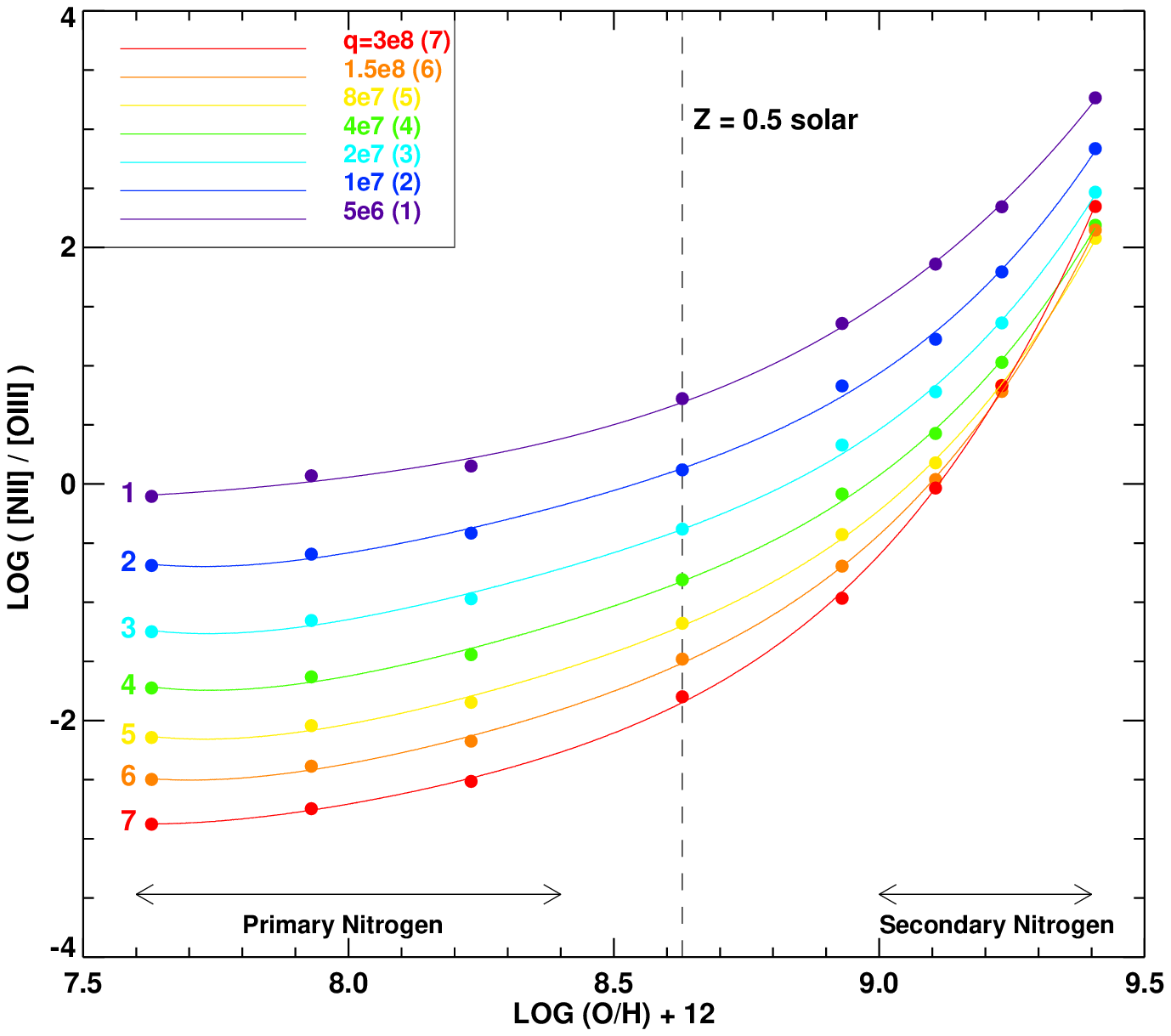}
\caption[f8.eps]{The log([\ion{N}{2}]~$\lambda 6584$/[\ion{O}{3}]~$\lambda 5007$ diagnostic for abundance versus metallicity.  Curves
for each ionization parameter between ${\rm q} = 5\times10^{6}$ to
$3\times10^{8}$ cm/s
are shown.  Filled circles represent the data points from our
models at metallicities from left to right of 0.05, 0.1, 0.2,
0.5, 1.0, 1.5, 2.0, 3.0 \Zsun. This figure is available in
color from the on-line version of this article.
\label{NIIOIII_vs_Z}}
\end{figure}

Our calibration of the \NIIOIII\ ratio is shown in
Figure~\ref{NIIOIII_vs_Z}.  The fourth-order polynomial fit coefficients
are given in Table~\ref{coefftable}.

\placefigure{NIIOIII_vs_Z}

Because the two ions have quite different ionization potentials, the
\NIIOIII\ ratio depends strongly on the ionization parameter.  Thus, if
this diagnostic is to be useful for abundance determinations, it must be
used in combination with an independant ionization parameter diagnostic.
However, if the \OII\ or \SII, and [\ion{S}{3}] lines are available
to determine the ionization parameter, then it would make more sense
to use either the \NIIOII\ or \NIISII\ abundance diagnostics rather than
this one.

\subsection{Abundance \& Ionization Parameter from these Line Ratios}

\begin{figure}
\plotone{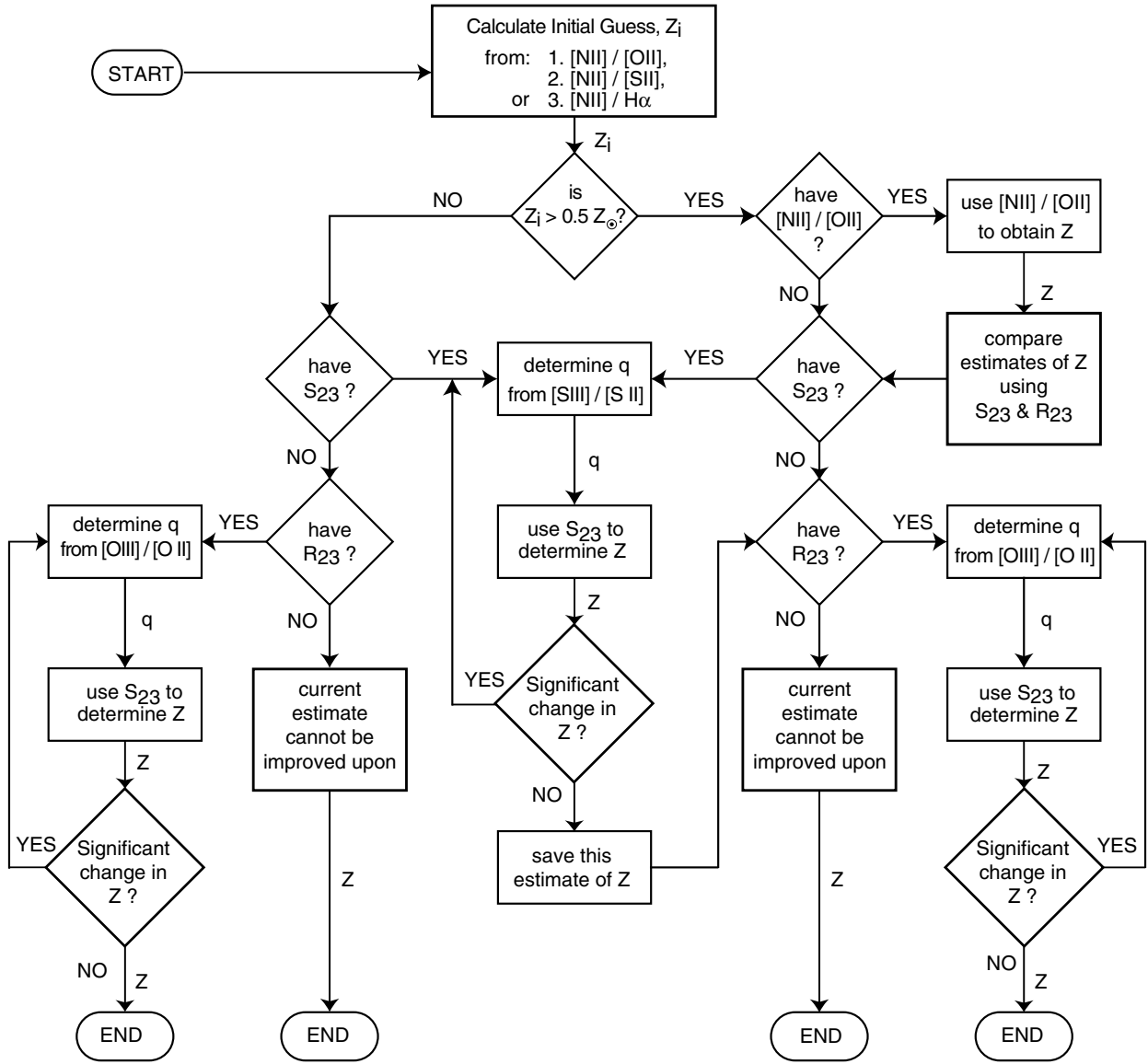}
\caption[f9.eps]{A logical flow diagram for determining abundance and
ionization parameter using optical emission-line diagnostics.  An IDL
script is available from the first author which automatically carries
out this process given a set of emission-line fluxes.\label{flow}}
\end{figure}

Since, in real data sets, we may have available only some of these line
ratios, we have summarised in Figure~\ref{flow} a logical process whereby
as many estimates of both the chemical abundance and ionization parameters
as are compatible with the data set can be obtained using the techniques
described in this section. The process can be automated, and an IDL script
to do this is available on request from the first author (LJK). However,
 we will describe a more direct method for
the derivation of these parameters, in Section \ref{abundance} below. 
From the comparative analysis with
other authors' techniques which follows, we are confident that this new
technique will provide the most reliable abundance estimates currently
possible.

\placefigure{flow}

\section{Comparison with other Bright-Line Techniques}
\label{comparison}

\subsection{Comparison Data}
\label{data}

Most of the data sets previously used to compare and calibrate abundance
diagnostics had been selected in different and heterogeneous ways: some by
galaxy (brightest \HII\ regions, or brightest disk \HII\ regions), some
by objective prism searches (which are biased towards strong 
\OIII~$\lambda \lambda$49595007),
some by Galaxy type (such as dwarf irregulars).  These different data sets
are reflected in the differences between the various calibrations of
abundance.  Care must be taken therefore when comparing different abundance
diagnostics to take into account biases (if any) introduced by the comparison
data.

We chose to use observations of \HII\ regions available from the large and
homogeneous data set of \citet{vanZee98}.  These authors observed 185 \HII\
regions in 13 spiral galaxies with the Double Spectrograph on the Palomar
5m telescope. These data have the additional advantage of covering a large
range in metallicity and ionization parameter, as was shown in 
\citet{Dopita00a}.

Since \SIII\ measurements are not available for the van Zee \HII\ regions,
we have also used two additional data sets for the \dS23 diagnostic;
\citet{Dennefeld83} and \citet{Kennicutt96}.  These also cover a wide range
in ionization parameter and metallicity.  Using the ESO 3.6m telescope,
\citet{Dennefeld83} observed $\sim 40$ \HII\ regions in the the 
Magellanic Clouds
and the Galaxy, while \citet{Kennicutt96} observed a similar number of \HII\
regions in M101 using the 2.1m telescope at Kitt Peak.

\subsection{Comparison Techniques for deriving Abundance}
\label{Comparison_techniques}

As discussed in the introduction, a wide number of empirical and
semi-empirical approaches already exist for the determination of abundances
in \HII\ regions.  We compare three of the most commonly-used  calibrations
produced by \citet{McGaugh91}, (hereafter M91), \citet{Zaritsky94}
(hereafter Z94), and \citet{Charlot01}(herafter C01) with the results produced
by our proposed theoretical diagnostics.

M91's calibration of \R23 makes use of detailed \HII\ region models
using the photoionization code CLOUDY \citep{Ferland81}, which also includes
the effects of dust and variations in ionization parameter.  We have used
the analytic expressions for the M91 models given in \citet{Kobulnicky99b}.
The Z94 calibration of \R23\ for the metal-rich regime is an average
of the three calibrations given by \citet{Edmunds84,Dopita86,McCall85},
with the uncertainty being estimated by the difference between the three
determinations.   A solution for the ionization parameter is not explicitly
included in the Z94 calibration.

C01 gives a number of calibrations depending on the availability of
observations of particular spectral lines. Their calibrations are based on
a combination of stellar population synthesis and photoionization codes with a
simple dust prescription, and include ratios to account for the ionization
parameter.  As the \HII\ regions in our sample contain measurements of
[\ion{O}{3}], \OII, \NII\, \SII\ and the Balmer lines, we use Case A of C01,
which is based on the \NIISII\ ratio, with a small dependance on \OIIIOII\
for ionization parameter correction.

\begin{figure}
\epsscale{0.5}
\plotone{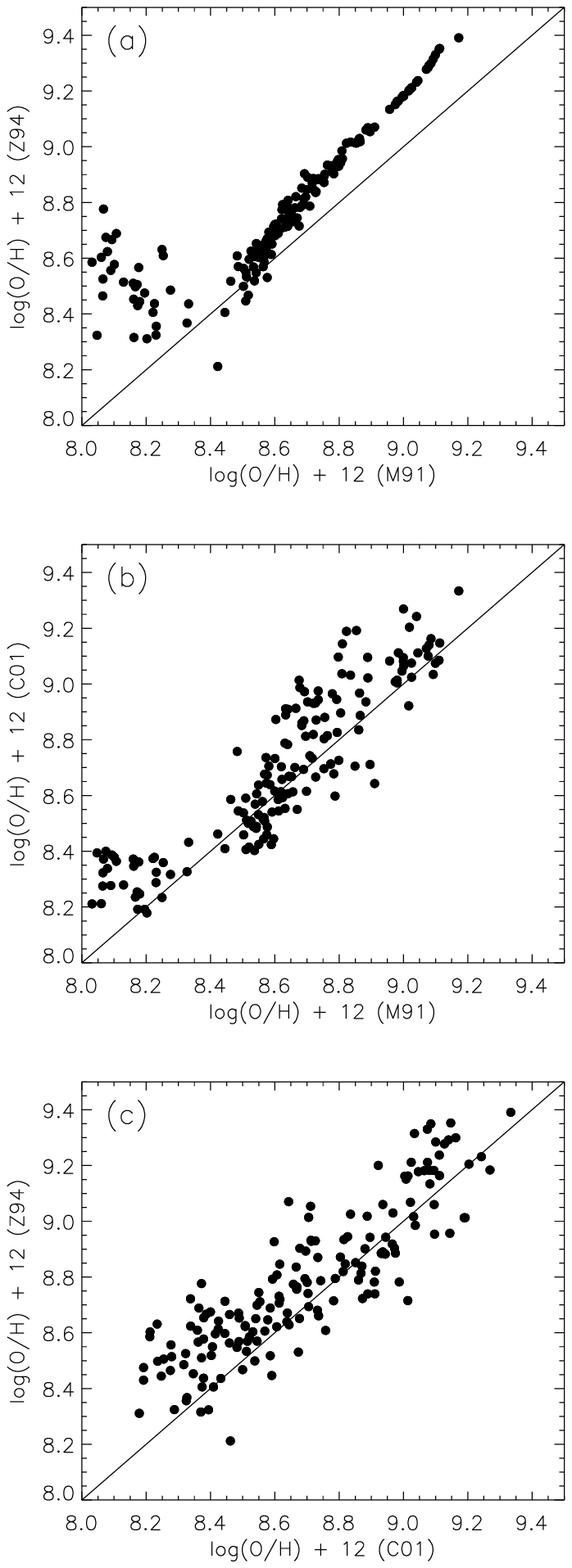}
\caption[f10.eps]{Oxygen abundance for the \citet{vanZee98} \HII\ regions 
using \citet{McGaugh91} (M91), \citet{Zaritsky94} (Z94) and
\citet{Charlot01} (C01).
\label{M91_vs_Z94_vs_C01}}
\end{figure}

Figure~\ref{M91_vs_Z94_vs_C01} shows the oxygen abundances obtained using
the Z94, M91 and C01 calibrations for the galaxies in our sample.
These clearly show the systematic and random errors associated with each
of these calibrations, but it is rather more difficult to estimate the
absolute reliability of any particular technique.

\placefigure{M91_vs_Z94_vs_C01}

The difference between the M91 and Z94 calibrations
(Figure~\ref{M91_vs_Z94_vs_C01}a)
is due to a number of factors.  The M91 calibration contains a correction
factor based on the (\OIII~$\lambda \lambda$49595007/ \OII~$\lambda$3727) ratio in order to account for
ionization parameter variations, while the Z94 calibration does not.  Z94's
diagnostic was calibrated against a large sample of disk \HII\ regions
which span the range \OH $> 8.35$.  As a result, their calibration
is only suitable for \HII\ regions in the metal-rich regime, as
noted by \citet{Kobulnicky99b}.  M91's diagnostic was calibrated against
\HII\ regions from a large range of sources from the literature, including
many \HII\ galaxies from the \citet{Campbell88} objective prism sample,
and is parametrized by one polynomial for the metal-poor regime and one
for the metal-rich regime.

The C01 calibration when compared with either the Z94 or M91 results shows a
large degree of scatter on these diagrams.  The C01 calibration (case A)
is based on \NIISII .  This ratio is not as sensitive to abundance as
\R23, and therefore introduces a large degree of scatter.  It is
also quite dependant on ionization parameter, which can also increase
the scatter.
Any technique based on \NIISII\ will have a similar amount of scatter, or
in other words, a
higher uncertainty in the abundance determined.  They are
therefore not as useful when comparing relative abundances as the
techniques with less uncertainty such as \R23\ or \NIIOII .

The problem of course when comparing abundances derived in this way, is
that there is no definitive `correct' answer, since we do not have
the \OIII\ Auroral line as a comparison.
In an attempt to overcome this we have taken the average of the
abundances derived using all three methods above as a `comparison' abundance.
Recalling that the Z94 method is actually based on three previous abundance
techniques, this `comparison' abundance is then the average of five virtually
independant methods.  A single abundance diagnostic will be described below
as `reliable' if it can produce the abundances similar to the comparison
abundance and also demonstrates a small degree of scatter.  The solid line
shown in the has the form $y=x$.  In the ideal case, with the
comparison abundance on one axis and the abundance estimate using a particular
technique on the other axis,
the data would lie along this solid line.  Therefore, the error in an abundance
technique can be represented by the root-mean-square distance (rms) of the data
to this line.   The rms error, given in Table~\ref{rms_table}, will naturally be
increased by both a systematic
shift and random scatter, so the rms error should be considered along with the
corresponding figure to understand why a particular technique is different
to the comparison abundance.  Ideally, the rms error should be equal 
to the rms introduced by uncertainties in the emission-line fluxes 
measured, which we estimate to be $0.02-0.03$ rms in $\log({\rm O/H})$ 
for the van Zee data set.
Note that an rms error of $\sim 0.1$ or more in $\log({\rm O/H})$ corresponds to an
rms error of $> 0.5 \times$\Zsun\ at high abundances (2$\times$\Zsun) and
$> 0.1 \times$\Zsun\ at low abundances (0.2$\times$\Zsun).  For
\HII\ regions or star forming galaxies, this level of accuracy is quite
adequate at low abundances, but it may not be sufficient at high abundances.
An rms error close to $0.03$ in $\log({\rm O/H})$ ($\sim 0.1 \times$\Zsun) would be 
would be desirable at high abundances. This is especially important in, 
for example,
studies of galactic abundance gradients using \HII\ regions.

\begin{figure}
\plotone{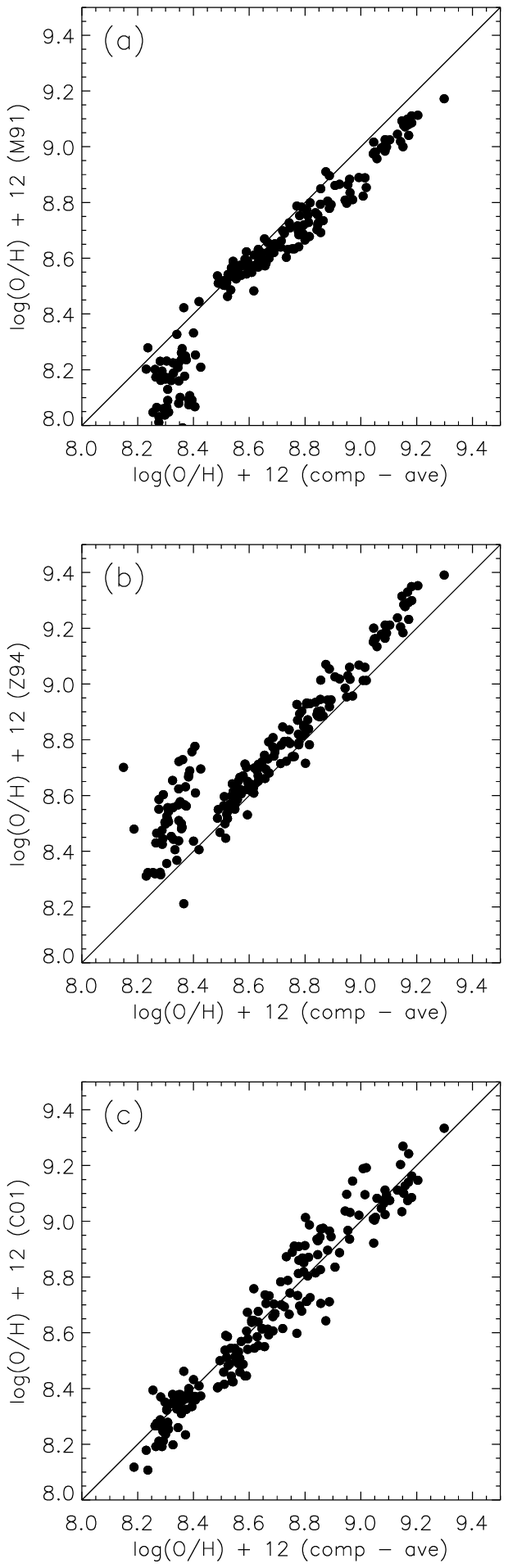}
\caption[f11.eps]{Oxygen abundance for the \citet{vanZee98} \HII\ regions 
using \citet{McGaugh91} (M91), \citet{Zaritsky94} (Z94) and
\citet{Charlot01} (C01) compared with the average of these three
methods (comp - ave).
\label{M91_Z94_C01_vs_ave}}
\end{figure}

First let us see how each of the three comparison techniques; M91, Z94,
and C01 compare with the average of all three, shown in
Figure~\ref{M91_Z94_C01_vs_ave}.

\placefigure{M91_Z94_C01_vs_ave}

Care must be taken in the interpretation of the panels of
Figure~\ref{M91_Z94_C01_vs_ave}, since each technique used on the y-axis
also contributes to the global average used on the x-axis.  This
tends to reduce any systematic shift and also decreases the random error.
The rms scatter and systematic offsets estimated for any of these techniques are therefore lower limits

\begin{figure}
\plotone{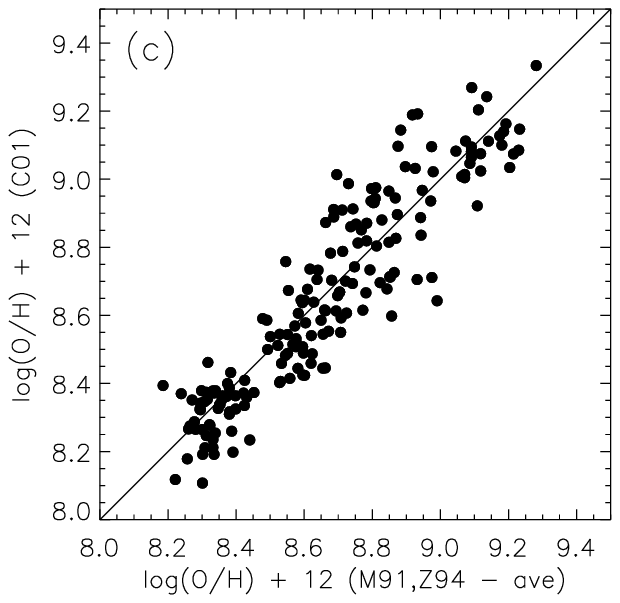}
\vspace{-12cm}
\caption[f12.eps]{Oxygen abundance for the \citet{vanZee98} \HII\ 
regions found with \citet{Charlot01} (C01) compared 
with the average of
the abundance from the \citet{McGaugh91} (M91) and \citet{Zaritsky94} methods.
\label{M91Z94ave_vs_C01}}
\end{figure}

With these caveats, it is clear that the M91 method delivers a systematically
lower abundance estimate than the average, and becomes quite unreliable
for abundances \OH $< 8.5$.
The M91 technique is based on \R23\, and as
we have already seen, \R23\ becomes less sensitive to abundance in this
regime.  The Z94 method tends to give a higher abundance than the average,
and is also not reliable for abundances \OH $< 8.5$ for the same reason.
The observation that M91 and Z94 are not reliable for abundances 
\OH $< 8.5$ has also been made by \citet{vanZee98} who found a similar
problem with the \R23\ diagnostic of \citet{Edmunds84}.
The C01 estimates appear to be reliable for all abundances, but is based
upon the \NIISII\ method, which has a high degree of intrinsic scatter.
This scatter is much reduced in this diagram because the C01
abundance estimate is also contributing significantly to the average here.
This is verified by Figure~\ref{M91Z94ave_vs_C01}, which compares the
average of M91 and Z94 abundances with those of C01. Since the M91 and
Z94 methods are tightly correlated (Figure~\ref{M91_vs_Z94_vs_C01}a),
the majority of the rms scatter of 0.09 arises with the C01 abundance
estimates.  Note that since there is no systematic shift observed in
this diagram, the average of all three diagnostics should have little
or no systematic errors, unless all three are subject to exactly the 
same form of systematic errors. The latter is a highly unlikely possibility 
given that all three methods are independantly derived, based on different line
ratios, they use different models, and they are calibrated against
different data sets.  This gives us confidence that, within the scatter,
the comparison abundance provides a reliable indicator of the true
abundance of a particular \HII\ region or star forming galaxy.

\placefigure{M91Z94ave_vs_C01}

We now proceed to compare our techniques with the average of the comparison
techniques.  Note that since we have shown that \NIIHa\ is relatively
insensitive to abundance and is good for an initial guess only, it will not
be used further in the analysis.

\subsection{The \NIIOII\ Diagnostic}
\label{NIIOIIdiagnostic}

\begin{figure}
\plotone{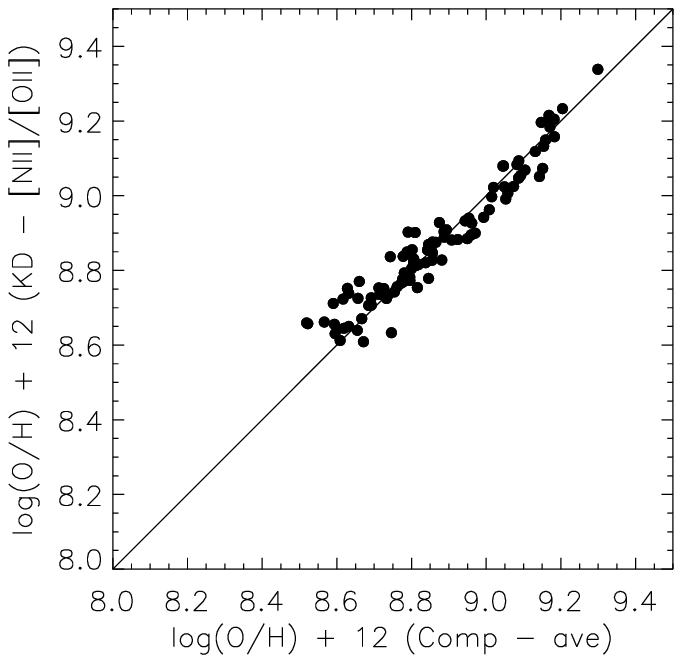}
\caption[f13.eps]{Oxygen abundance for the \citet{vanZee98} 
\HII\ 
regions found with our (KD) \NIIOII\ diagnostic compared with the average (comp - ave) of
those found using
\citet{McGaugh91}, \citet{Zaritsky94} and
\citet{Charlot01}.
\label{KNIIOII_ave}}
\end{figure}

Our estimate of abundance using \NIIOII\ (Figure~\ref{NIIOII_vs_Z}) is
shown in Figure~\ref{KNIIOII_ave}.

\placefigure{KNIIOII_ave}

Clearly, this provides a very reliable abundance diagnostic.  This diagram
displays the smallest scatter of any of the techniques presented so far,
with an rms error of 0.04.
Such a strong correlation between our \NIIOII\ technique and the average of
five previous techniques gives confidence not only in the derived abundance,
but also in the reddening correction which has been applied to the observed
spectra using the `classical' reddening curves. Clearly, the Balmer
decrement has been reliably measured in the \HII\ regions of the
van Zee sample.

To better estimate the effect of reddened spectra on the \NIIOII\
abundance determination, a random reddening corresponding to E(B-V)
values between 0 and 1.0 was added to the dereddened fluxes measured in
the van Zee \HII\ region spectra.  This simulates observations of a sample
of \HII\ regions with an average E(B-V) of 0.5.

\begin{figure}
\plotone{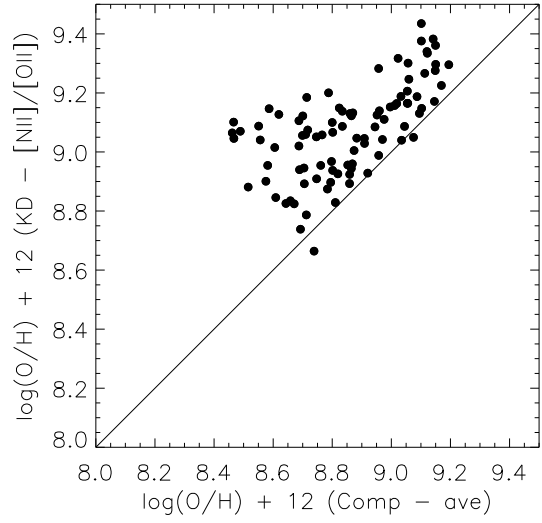}
\caption[f14.eps]{As in Figure~\ref{KNIIOII_ave}, but
for \HII\ regions with random reddening of $0.0 \le {\rm E(B-V)} \le 1.0$
added to their spectra.
\label{KNIIOII_ave_extinct}}
\end{figure}

If we were to then use the \NIIOII\ diagnostic on this ``reddened" sample
without correction for reddening, then the abundances estimated are
much more scattered, with a systematic shift to higher derived abundances,
as shown in Figure~\ref{KNIIOII_ave_extinct}.  The rms error is now 0.14,
and a systematic shift of \OH~$=0.5$ is observed.

\placefigure{KNIIOII_ave_extinct}

It is clear that, when the Balmer decrement (or another means to
estimate E(B-V)) is unavailable, and reddening is believed to be
important, the \NIIOII\ method should not be used.

Note that for large samples of \HII\ regions or starburst galaxies,
the tightness of the \NIIOII\ versus the comparison average may be a
good means to test the reliability of the E(B-V) determination.  If a
shift in this abundance estimate is observed relative to other indicators,
then E(B-V) is being systematically underestimated, and if there is a
scatter greater than 0.04 rms, then the estimated reddening values should
be treated with caution.  A random effect such as this might
be observed for example in a sample where differing dust geometry affects
the reddening of each object in a different fashion.

\subsection{The $\mathbf{{\rm R}_{23}}$ Diagnostic
\label{R23diagnostic}}

\begin{figure}
\plotone{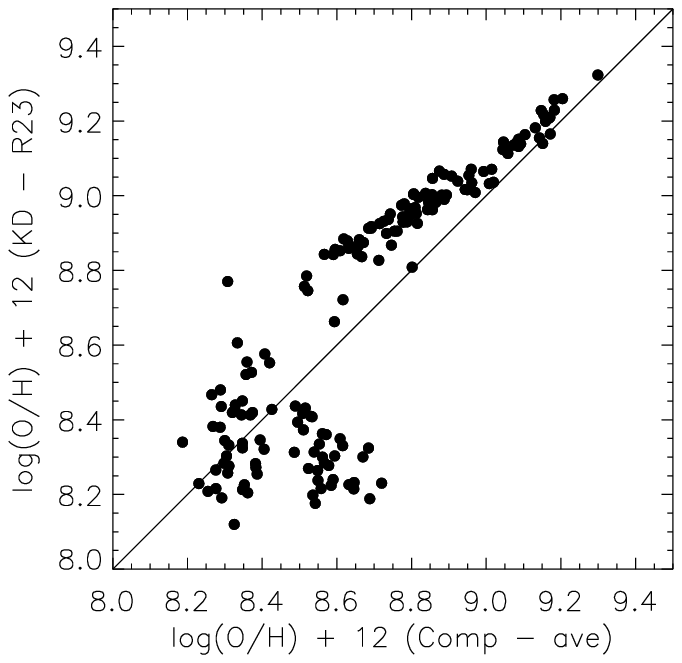}
\caption[f15.eps]{Oxygen abundance for the \citet{vanZee98} \HII\ 
regions 
found with our (KD) \R23\ diagnostic compared with abundances found
from the average (comp - ave) of
\citet{McGaugh91}, \citet{Zaritsky94} and
\citet{Charlot01}.
\label{KR23_ave}}
\end{figure}

For the \R23\ diagnostoc, an initial guess from \NIIOII\
was used, and the ionization parameter
determined using \OIIIOII.  The abundance was then determined
using \R23, with the branch of the \R23 curve chosen depending on
whether the initial guess was greater or less than the \R23 maximum
(\OH~$=8.5$).
This process was iterated with \OIIIOII\ until the abundance did not vary.
For 88\% of the 185 galaxies in the van Zee sample, the abundance converged
after 1 iteration.  Results for \R23 compared with the average of
M91,Z94 and C01 are shown in Figure~\ref{KR23_ave}.

\placefigure{KR23_ave}

Abundances in the region $8.5 \le $\OH$ \le 8.9$ cannot be reliably
determined due to the maximum in the \R23\ versus abundance curve.
We see two systematic effects operating in this abundance range.
Firstly, there is a systematic shift towards higher
estimated abundances than the comparison abundance, which is similar to
shifts seen with other \R23\ diagnostics (eg. Figure~\ref{M91_vs_Z94_vs_C01}a).
This is a result of the rapidly declining sensitivity to abundance as
the \R23\ curve reaches its local maximum at around \OH~$=8.5$.  The other
systematic effect observed is  for abundances between 8.5 and 8.8,
where the \R23\ diagnostic estimates significantly lower abundances
between 8.2 and 8.4, causing a branch to appear almost perpendicular to
the $y=x$ solid line.  This is also a result of the lack of sensitivity
of the \R23\ diagnostic around its local maximum.  If the comparison abundance
is correct (and we believe that it is), then these \HII\ regions mostly have an
abundance of around \OH~$=8.5-8.6$, which is exactly at the local maximum of
\R23.  An \HII\ region with an abundance of \OH~$=8.5-8.6$ will produce
\R23 abundance estimates anywhere in the range $\sim 8.8$ or $\sim 8.3$,
since this diagnostic is almost insensitive to metallicity throughout
this range.

For \OH $\ge 8.9$, our \R23\ method estimates a somewhat higher abundance
estimate than the comparison abundance, but this model-dependent shift is 
no greater than that seen for other \R23\ -based diagnostics (M91 and Z94)
in this region.  For \OH$ \le 8.5$, our \R23\ method shows no systematic
shift, and gives an rms scatter of 0.07 , which for this abundance range
corresponds to an rms of $\sim 0.1 \times$\Zsun.  This is pleasing, when we
consider that some other \R23\ techniques fail in this region
(Figure~\ref{M91_Z94_C01_vs_ave}a,b).
The scatter we obtain is similar to that obtained in the C01 \NIISII\
method.

\subsection{The \NIISII\ Diagnostic
\label{NIISIIdiagnostic}}

\begin{figure}
\plotone{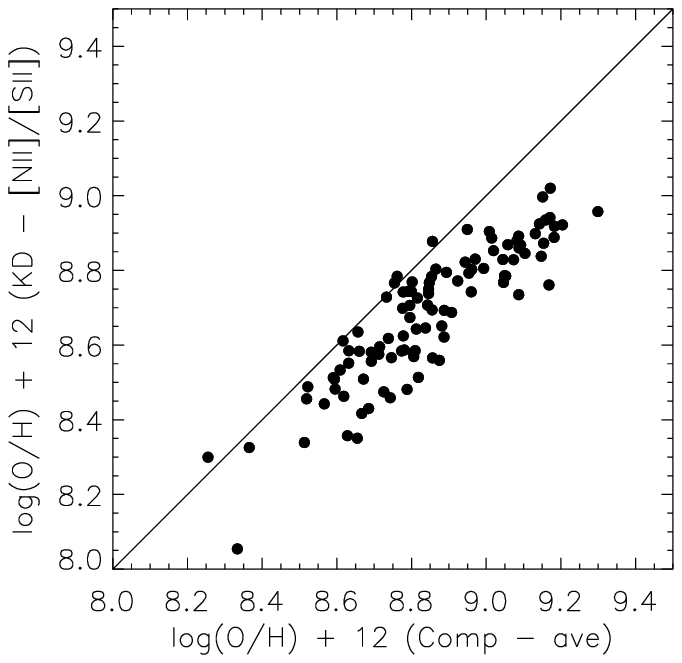}
\caption[f16.eps]{Oxygen abundance for the \citet{vanZee98} 
\HII\ regions 
found with our (KD) \NIISII\ diagnostic compared with abundances found
from the average (comp - ave) of
\citet{McGaugh91}, \citet{Zaritsky94} and
\citet{Charlot01}.
\label{KNIISII_ave}}
\end{figure}

Since the \NIISII\ diagnostic is primarily sensitive in the high metallicity
range and is also sensitive to ionization parameter in this regime, the
preferred method is to use an ionization parameter diagnostic such as
\OIIIOII\ .  However, if \OII\ is available, then it would make more
sense to use the \NIIOII\ diagnostic to derive a more reliable abundance,
since \NIIOII\ is much more sensitive to abundance than the \NIISII\ ratio.
Nevertheless, for completeness, Figure~\ref{KNIISII_ave} compares the
abundance obtained with our \NIISII\ method with the average of the
three comparison techniques.

\placefigure{KNIISII_ave}

Figure~\ref{KNIISII_ave} shows a large degree of
scatter, similar to that seen in the C01 scheme, which is also based on
the \NIISII\ ratio.  This scatter is inherent to using the \NIISII\ ratio
which is not as sensitive to abundance as other ratios, particularly in
the lower metallicity range.  In addition, \SII\ is affected by the electron
density of the line-emitting regions, and this may also contribute towards
the scatter. There is also a systematic shift in our estimate of abundance,
in the sense that our \NIISII\ technique will systematically underestimate
the abundance by a factor of $\sim 0.2$ dex compared to the average of the
comparison techniques.  We believe this is a result either of the use of
a different sulfur to oxygen abundance ratio or the depletion
factors used in our models for \SII. Both the large scatter and
systematic shift result in a large rms of 0.18 .

The sulfur to oxygen abundance ratio has been found to be the least
accurately determined amongst the elements which are usually
observed in planetary nebulae \citep[eg.][]{Natta80}.  The determination of
the sulfur abundance is particularly difficult because of the large
ionization correction factors that must be applied to correct for the
presence of unobserved ionization states.  These ionization correction
factors are often made based on a procedure developed by \citet{Peimbert69},
which assumes that ions with similar ionization potentials
are equally populated.  This procedure has been shown to overestimate
the S$^{+3}$ abundance by a factor of more than three for most \HII\ regions
\citep{Natta80}.  While it is possible to obtain reliable S/O ratios for
some low and moderate excitation \HII\ regions \citep{Dennefeld83}, it is
unreliable for high excitation \HII\ regions \citep{Garnett89}.  Furthermore,
Garnett showed that the [\ion{S}{2}]/[\ion{S}{3}] ratio is
underpredicted by ionization models which produce [\ion{O}{2}]/[\ion{O}{3}]
ratios comparable to observations.  Garnett suggests this is a result of
uncertainties in model stellar atmosphere fluxes or the atomic data for
sulfur.  In the light of this discussion, we can probably conclude that
the systematic offset that we observe for \NIISII\ reflects the systematic
errors inherent in any abundance or ionization parameter diagnostic
involving sulfur.

\subsection{The \dS23 Diagnostic
\label{S23diagnostic}}

\begin{figure}
\plotone{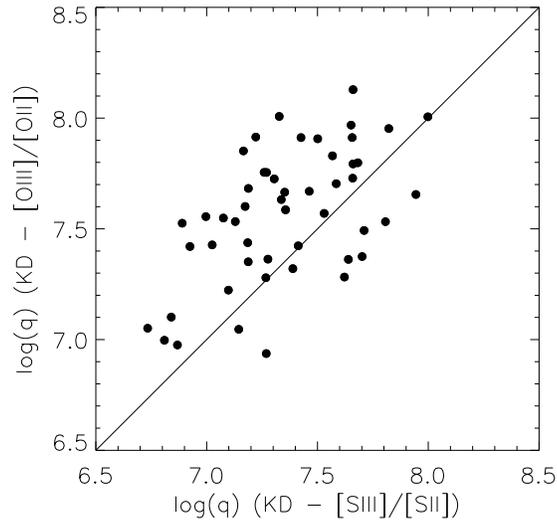}
\caption[f17.eps]{Ionization parameter found with our (KD) \SIIISII\
diagnostic versus ionization parameter found with our \OIIIOII\ diagnostic
for the \citet{Kennicutt96} and \citet{Dennefeld83} \HII\ regions.
Ionization parameter is in units of cm/s.
\label{qOIIIOII_vs_qSIIISII}}
\end{figure}

Despite the problems with modeling sulfur as discussed above, and our
suspicion that the \dS23\ diagnostic may not provide a reliable abundance
estimate, we have nonetheless applied the \dS23\ ratio to the \HII\
region data from \citet{Dennefeld83} and \citet{Kennicutt96} and derived
abundances.  The ionization parameter was first estimated using the
\SIIISII\ ratio.  The ionization parameter obtained with this ratio
is compared with that found with the \OIIIOII\ ratio in
Figure~\ref{qOIIIOII_vs_qSIIISII}.

\placefigure{qOIIIOII_vs_qSIIISII}

This figure shows a large scatter between the two techniques (rms=0.15).
Such a scatter in determinining the ionization parameter could introduce
an rms scatter in abundance of between 0-0.2 in \OH, depending on which
diagnostic is used.  If the error is concentrated mainly in the \SIIISII\
ratio, then this alone would be sufficient to cause the large scatter
obtained for the \NIISII\ diagnostic.

Figure~\ref{qOIIIOII_vs_qSIIISII} also shows that the \SIIISII\ ratio gives
an ionization parameter which is systematically $\sim 0.2$~dex smaller than
that obtained from \OIIIOII\ .  This means that all of the \HII\ regions
will have an estimated $q < 1.5 \times 10^{8}$~cm/s, the regime for which 
the \dS23\
curve is actually not very sensitive to ionization parameter.

\begin{figure}
\plotone{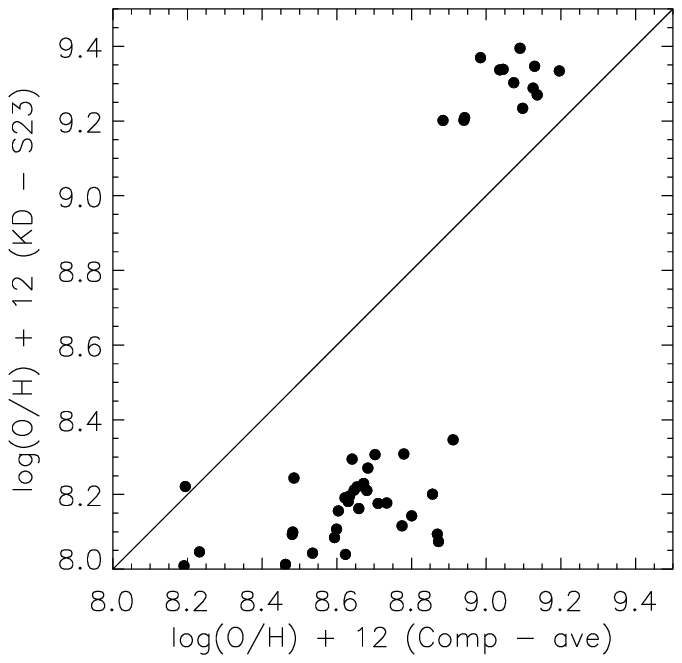}
\caption[f18.eps]{Oxygen abundance for the \citet{vanZee98} 
\HII\ regions 
found with our (KD) \dS23 diagnostic compared with abundances found
from the average (comp - ave) of
\citet{McGaugh91}, \citet{Zaritsky94} and
\citet{Charlot01}.
\label{S23_vs_compave}}
\end{figure}

The abundance derived using our \dS23\ diagnostic is compared with the
comparison abundance in Figure~\ref{S23_vs_compave}.  Clearly this
diagnostic does not reliably estimate abundances for any metallicity
range, and should therefore not be used.  A reliable theoretical
\dS23\ diagnostic will have to await a resolution of the uncertainties
associated with sulfur modeling.

\placefigure{S23_vs_compave}

The fact that \dS23\ underestimates the abundance at low metallicities
and overestimates at high metallicities means that the location of the
peak of our \dS23\ grid is too high.  Both this and the systematic shift
in abundance obtained for the \NIISII\ and \SIIISII\ line ratios can be
explained if the sulfur abundance in our models has been under-estimated
and that the \SIIISII\ ratio has been over-estimated.  That the
\SIIISII\ ratio has been over-estimated in abundance modelling has been
previously suggested by \citet{Garnett89} as discussed above.

\begin{figure}
\plotone{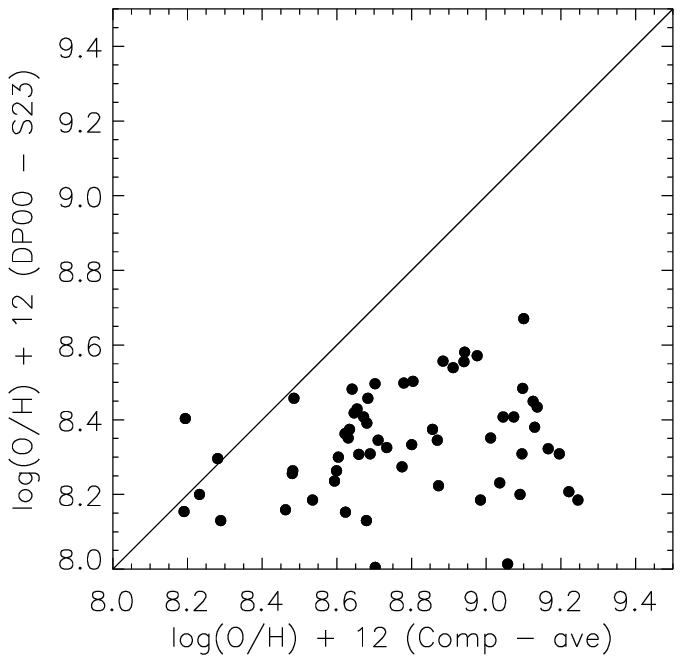}
\caption[f19.eps]{Oxygen abundance 
found with the \citet{Diaz00} (DP00) diagnostic compared with abundances found
from the average (comp - ave) of
\citet{McGaugh91}, \citet{Zaritsky94} and
\citet{Charlot01} for the \citet{Kennicutt96} and 
\citet{Dennefeld83} \HII\ regions.
\label{OtherAve_vs_DiazS23}}
\end{figure}

Recently, \citet{Diaz00} (hereafter DP00) suggested the use of an
empirical \dS23\ diagnostic which they tested using \HII\ regions for which
direct determinations for electron temperature were available. These
include the \citet{Dennefeld83} \HII\ regions.  The DP00 abundance
estimate is compared with the comparison abundance in
Figure~\ref{OtherAve_vs_DiazS23}.

\placefigure{OtherAve_vs_DiazS23}

The DP00 diagnostic systematically underestimates the abundance
relative to the comparison abundance.  This underestimation becomes
progressively worse at higher metallicity. There is also a very large
scatter (rms=0.39). This is probably a result of the linear calibration
used by DP00. From Figure~\ref{S23_vs_Z}, it is clear that \dS23\ can
only be fit by a linear calibration up to abundances of about
\OH $=8.4$, after which the curve begins to turn over.
It is interesting that the systematic shift observed here is similar
to that found by \citet{Kobulnicky99b}, in that for low metallicity
galaxies the \O4363\ diagnostic systematically underestimated the global
oxygen abundance.  Nevertheless, in the absence of reliable
theoretical \dS23\ models, empirical calibrations such as that by
\citet{Diaz00} give hope that \dS23\ can be used as an abundance
diagnostic.  Indeed, if we add an extra term to the DP00
calibration, such as

\begin{equation}
\log({\rm O/H}) + 12 = 1.53\,{\rm S}_{23} +8.27 + 
\frac{1.0}{2.0-9.0\,({\rm S}_{23})^{3}}
\label{eq_DiazS23_extraterm}
\end{equation}

\begin{figure}
\plotone{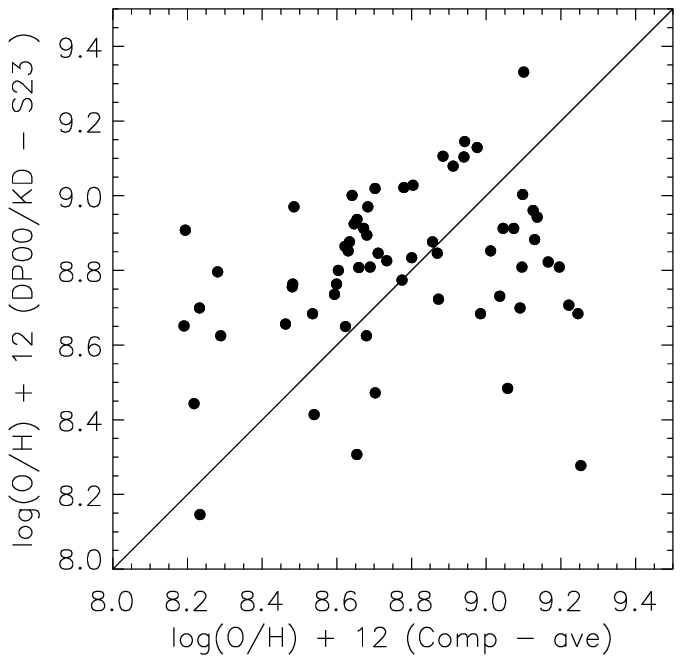}
\caption[f20.eps]{Oxygen abundance 
found with the \citet{Diaz00} diagnostic with an additional empirical 
term (DP00/KD) added
compared with abundances found from the average (comp - ave) of
\citet{McGaugh91}, \citet{Zaritsky94} and
\citet{Charlot01} for the \citet{Kennicutt96} and 
\citet{Dennefeld83} \HII\ regions.
\label{OtherAve_vs_DiazS23_2}}
\end{figure}

(shown in Figure~\ref{OtherAve_vs_DiazS23_2}),
we can produce a correlation between the comparison abundance
and the abundance derived above, with a large but significantly reduced
rms error of 0.22.

Clearly, more work needs to be done before we can derive a reliable
\dS23 diagnostic.

\placefigure{OtherAve_vs_DiazS23_2}

\section{Optimized abundance determination: Recommended method
\label{Combined_method}}

As we have seen, with the exception of our \NIIOII\ diagnostic, all of
the comparison techniques and our \R23, \NIISII, and \dS23\ schemes
are plagued by systematic and/or random errors.  Nonetheless, some of
these techniques, individually or in combination, are reliable
over limited ranges of metallicity. It is therefore possible to derive a
combined technique which can be used over the entire range of abundances
from \OH$=8.2$ up to \OH$=9.4$.  Ideally, this technique will have a small
rms, similar to that found with our \NIIOII\ method, and display no
systematic shift relative to the comparison abundance.

The technique presented here applies these criteria, and is applicable
to objects for which the spectra include a range of bright emission lines,
such as \NII, \OII, \OIII, and \SII.

First, we have shown that our \NIIOII\ diagnostic provides an
excellent means of determining abundance for metallicities above \OH$=8.6$.
For this abundance range, the relationship between the \NIIOII\ ratio
and abundance can be fit by a simple quadratic:

\begin{equation}
\log({\rm O/H}) + 12 =  \log[1.54020+1.26602\,{\rm R}
                             +0.167977\,{\rm R}^2] + 8.93
\end{equation}

where R is $\log($\NIIOII). This applies for the high metallicity regime
[$\log({\rm O/H}) +12 \ge 8.6$], and in this regime, the abundance derived
from the above equation should be taken as a final abundance estimate.
If required, the ionization parameter appropriate to the region observed
can then be determined using Figure~\ref{OIIIOII_vs_q} 
(equation~\ref{eq_OIIIOII}).

Next, if application of the above equation yields an abundance below 8.6,
then the average of the M91 and the Z94 methods can be used to provide an
abundance estimate.  The equation for the Z94 \citep{Zaritsky94} method and
the M91 \citep{McGaugh91} method as parametrized by \citet{Kobulnicky99b}
are provided below for reference.  Note that Kobulnicky, Kennicutt \& 
Pizagno provided two equations for M91, one for the lower metallicity
branch (\OH$\le 8.4$) and one for the upper metallicity branch (\OH$> 8.4$)
The upper branch should be used here, the equation for which is given
below.

\begin{eqnarray}
{\rm Z94}:\, \log({\rm O/H}) + 12 & =&   9.265-0.33\,{\rm R}_{23}-
                       0.202\,{\rm R}_{23}^{2}-
                       0.207\,{\rm R}_{23}^{3}-
                       0.333\,{\rm R}_{23}^{4}   \\
{\rm M91(u)}: \, \log({\rm O/H}) + 12  &  = & 12.0 - 4.944 + 
0.767\,{\rm R}_{23}+0.602\,
{\rm R}^{2}_{23}- \nonumber \\
 & &y\,(0.29+0.332\,{\rm R}_{23}-0.331\,{\rm R}^{2}_{23})
\end{eqnarray}

where 

\begin{equation}
y=\log\left(\frac{[{\rm OIII}]\,\lambda \lambda 4959,5007}{[{\rm OII}]\,\lambda 3727}\right),
\end{equation}

unlike the \OIIIOII\ used in our ionization parameter diagnostic or the
C01 diagnostic, which are based on the \OIII~$\lambda$5007 and 
\OII~$\lambda$3727 lines only.

The average of these two diagnostics produces reliable abundances
down to a \OH\ of 8.5.

Finally, if this diagnostic gives an estimate below 8.5, then our \R23
method can be used.  There is a large degree of scatter inherent at this
abundance range for our \R23 method, very similar to that in the C01
method.  This scatter is reduced considerably by taking the average of
the two techniques.  The abundance should first be estimated using the
C01 method, as in the following equation, taken from from C01
\citep{Charlot01}:

\begin{equation}
\log({\rm O/H}) + 12= \log \left\{5.09 \times 10^{-4} \times \left[\left(\frac{{\rm 
[OII]/[OIII]}}{1.5}\right)^{0.17}\right]
\left[\left(\frac{{\rm [NII]/[SII]}}{0.85}\right)^{1.17}\right] \right\}+12
\end{equation}

Note that the line ratios are not used in the logarithmic form for the
application of the C01 method, as they are for all the other equations
given in this section. This abundance can now be used as an initial guess
as input to estimate the ionization parameter to be used in
our \R23\ method.

For abundances in this regime (below 8.5), the ionization parameter models 
for $R=\log($\OIIIOII) (Figure~\ref{OIIIOII_vs_q})
can be fit by a simple equation of the form:

\begin{equation}
q=10^{({k_{0} +k_{1}*{\rm R}+k_{2}*{\rm R}^{2}})}  \label{qOIIIOII_eq2}
\end{equation}

where $k_{0-2}$ are constants which depend on the initial guess provided
by the C01 method, and are given in Table~\ref{qcoefftable}.

Now, with a knowledge of the ionization parameter, we can estimate the
abundance using our \R23 method. In this abundance regime the lower
branch provides the appropriate solution, given by;

\begin{equation}
\log({\rm O/H}) + 12 = \frac{-k_{1} + \sqrt{k_{1}^{2} - 4 k_{2} (k_{0} -
                         {\rm R}_{23})}}{2 k_{2}}  \label{R23_eq2}
\end{equation}

where  $k_{0-2}$ are constants which depend on the ionization
parameter derived above, and are given in Table~\ref{coefftable}.

The solution obtained by the application of this equation can be then
substituted in equation \ref{qOIIIOII_eq2} to solve again for the
the ionization parameter, and the metallicity and the process can be
iterated until the solution converges.  Generally, we find that
convergence is attained after the first iteration in 88\% of cases.

The average of our \R23\ diagnostic and that given by the C01 method
is then adopted for abundances less than \OH$=8.5$.

\begin{figure}
\plotone{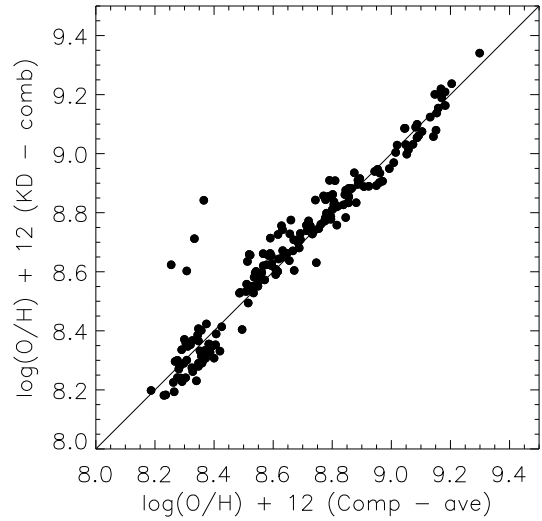}
\caption[f21.eps]{Oxygen abundance for the \citet{vanZee98} 
\HII\ regions 
found with our (KD) combined diagnostic compared with abundances found
from the comparison sample which is the average of
\citet{McGaugh91}, \citet{Zaritsky94} and
\citet{Charlot01}.  This combined technique minimizes the scatter
inherent
in other techniques and shows no systematic shift compared with the
comparison abundance.
\label{Combined_vs_otherave}}
\end{figure}

The results of this combined technique is shown against the comparison
abundance in Figure~\ref{Combined_vs_otherave}, and has an rms of 0.05.
With the exception of the \NIIOII\ diagnostic, the combined method
outlined above minimizes the scatter inherent in any of the methods
presented in this paper, either given by us or by C01, M91, and Z94.
In addition, it eliminates the systematic errors which characterize
many of the techniques when used alone.

There are four observed points out of 185 which lie off the tight
correlation in this figure.  These galaxies all have derived abundances
based on the \NIIOII\ ratio which are greater than 8.6, but have
comparison abundances between 8.2 and 8.4.  We have no reason not to
believe our \NIIOII\ abundance, especially when two of the three
comparison abundance diagnostics between 8.2-8.4 (M91 and Z94) are
not reliable (eg. Figure~\ref{M91_vs_Z94_vs_C01}a,b).  Without these
four outliers, the rms scatter would reduce still further, to 0.04.

There is a clear gap at abundance \OH~$ \sim 8.45$. This is a result
of the use of \R23\ diagnostics, which have a local maximum around
this point.  Objects which have a real abundance of 8.45 will typically
have an abundance estimated from the \R23\ ratio of either 8.4 or 8.5,
Relative to other errors, this is probably insignificant.

This diagram shows that with this procedure, and providing that we have
spectra which include observations of \OII, \NII, \OIII, \SII, and, of
course, the Balmer lines, it is possible to derive reliable abundances
over the whole range of abundances. This procedure utilizes the
strengths of various methods, while at the same time minimizes their
weaknesses.

\section{Solving for Abundances with Limited Data Sets}
\label{abundance}

As we have seen, due to the nature of the diagnostics, the excitation
potentials of the lines involved, and their dependance on ionization
parameter, different diagnostics are useful over different ranges of
metallicity.  Most line ratios display a local maximum, causing the
metallicity determination to be double valued unless another line
ratio can be used to determine on which branch the appropriate solution
lies. Others are dependant on ionization parameter over some metallicity
ranges, and if these are used, one needs first to solve for the ionization
parameter. In many cases, spectra may not cover the full wavelength
range needed for the application of the scheme recommended in the
previous section. Here we summarize procedures that should 
provide the most reliable abundance, given such a limited set of
emission-line fluxes.

\noindent
{\bf 1. \OII, \Hb, \OIII, \NII, \SII:}

If all of the above lines are available, then our combined method in
Section~\ref{Combined_method} should be used.  However, if \NIIOII\
cannot be corrected for reddening, and the reddening is expected to be
significant, then the average of M91, Z94 and C01 will provide
reasonable estimate for metallicities above 0.5$\times$solar.

{\bf 2. \OII, \OIII, \Hb, \NII}

In the absence of \SII, our combined method can still be used, but
for \OH~$<8.5$, our \R23\ method can be used alone rather than
averaging its results with the C01 method.  This will increase the
rms error in this abundance range to 0.07.

{\bf 3. \OII, \OIII, \Hb}

If only \OII, \OIII, and \Hb\  are available, then the \R23\
diagnostics must be used.  Since all the \R23\ methods we have looked
at contain systematic shifts which apply over particular abundance ranges,
the average of the metallicities estimated by a number of \R23\ methods
is desirable. Since the Z94 \R23\ method requires no initial guess,
this should be used as an initial guess for both our \R23\ method and
the M91 method to determine the appropriate solution branch on the \R23\
curve.  Note that the initial estimate based on Z94 will be very
inaccurate at low abundances (\OH~$<8.5$), causing metallicities
to be over-estimated by a factor of up to 0.5 in \OH\ in this range.
Therefore, in cases where the second \OH\ estimate (as described
below) is lower than the range shown, the abundance should be re-calculated 
using the method outlined below for the correct abundance range.

We then suggest the application of one of three cases based on this
initial guess:

(a) \OH~$> 9.0$:  First find the ionization parameter $q$ from
\OIIIOII\ using the Z94 estimate.  Next, use this ionization parameter
to find which coefficients to use for our \R23\ abundance estimate.
Estimate $q$ again and check whether it is sufficiently different
from the previous estimate of $q$ to influence the abundance derived.
Iterate until the abundance estimate converges.  Finally, the
abundance found with our \R23\ method should be averaged with those derived
from M91 and Z94.

(b) $8.5<$\OH~$\le 9.0$: The abundances from M91 and Z94 should be
averaged to obtain the abundance estimate in this range.

(c) \OH~$\le 8.5$: Our \R23 method alone should be used to determine
the abundance, since neither M91 nor Z94 give reliable
results in this range.  The ionization parameter should be found using
\OIIIOII\ and iterated with the abundance as described in (a).

\begin{figure}
\plotone{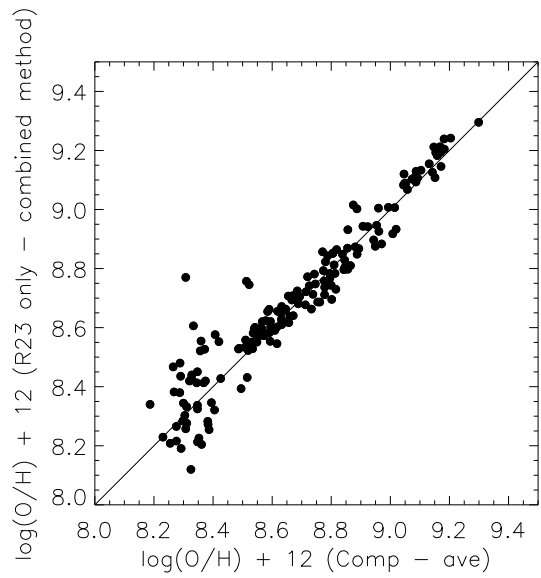}
\caption[f22.eps]{Oxygen abundance for the \citet{vanZee98} 
\HII\ regions 
found with our \R23 - only combined diagnostic compared with abundances found
from the comparison sample which is the average of
\citet{McGaugh91}, \citet{Zaritsky94} and
\citet{Charlot01}.  This combined technique minimizes the scatter
inherent
in other \R23 techniques used alone and shows no systematic shift 
compared with the
comparison abundance.
\label{OtherAve_vs_R23only}}
\end{figure}

We applied this method to the van Zee \HII\ regions and the results are
shown in Figure~\ref{OtherAve_vs_R23only}. This method displays an rms
scatter of 0.06.  Note however, that this is a lower limit because we
are using the Z94 and M91 methods for abundances greater than \OH~$=8.5$,
which also contribute towards the comparison average.

\placefigure{Combave_vs_ave_3}

{\bf 4. \NII, \OII:}

Suppose that only the red \NII\ and blue-UV \OII\ lines are available.
This circumstance might arise, for example, if we are dealing with an
intermediate-redshift galaxy ($1.2 \le z \le 1.7$) 
in which the \OII\ lines are observed in the red, but the
H$\alpha$ and \NII\ lines are observed in the IR. In this case, the
\NIIOII\ ratio should be used (Figure~\ref{NIIOII_vs_Z}) with
equation~\ref{poly4} after correction for reddening.  Since this
diagnostic is relatively independant of ionization parameter for this
metallicity range, no further steps are required to obtain the
metallicity.  For ${\rm Z} < 0.5\,{\rm Z}_{\odot}$ (\OH~$\le 8.6$), \NIIOII\
is rather insensitive to abundance and the estimate will be affected
by a large random error of 0.11 rms.

{\bf 5. \Ha, \NII, \SII:}

If only red spectra which include H$\alpha$, \NII\ and \SII\ are
available, then the C01 \NIISII\ technique cannot be used since it
requires a correction factor based on \OIIIOII\ .
Our \NIISII\ method produces a systematic shift of 0.2 dex.  It would be
possible therefore to first use the \NIISII\ shifted in \OH\ by 0.2 dex
combined with \NIIHa\ for an assumed average ionization parameter of
$q=2\times 10^{7}~cm/s$. Neither of these methods is ideal and both
depend strongly on the ionization parameter, which cannot be solved for
explicitly, so that this method will only give a very "rough guess"
of the abundance which is probably not useful for detailed
abundance studies.

{\bf 6. \SII, [\ion{S}{3}] :}

If the sulfur lines are the only ones available, then an empirical
abundance diagnostic such as \citet{Diaz00} as revised here
(equation~\ref{eq_DiazS23_extraterm}) should be used.

\section{Conclusions}
\label{conclusion}

We have presented a range of diagnostics using current theoretical
models for determining the abundances and ionization parameters of
star forming regions.  The appropriate diagnostics to be used depends
on the wavelengths observed, and therefore on the availability of
particular emission line ratios.  Our diagnostics have been
compared with those of \citet{McGaugh91}, \citet{Zaritsky94} and
\citet{Charlot01}, and we arrive at the following conclusions;

\begin{enumerate}
\item Our \NIIOII\ diagnostic is clearly the best
diagnostic to use in the Z$>0.5$~\Zsun\ (\OH~$>8.6$) regime, as it
produces a remarkably tight correlation with the abundance determined
using the average of five previous techniques, none of which can
reproduce such a tight correlation when used individually.
The \NIIOII\ diagnostic produces this tight correlation as a result
of both its independance of ionization parameter and its strong
metallicity sensitivity.  We have investigated the
effect of poorly reddening-corrected or un-corrected spectra on the
derived abundance using this diagnostic.  We show that this adds both
a large degree of uncertainty in the abundance derived, and a
systematic bias to estimate higher than average abundances.
However, we have shown that `classic' extinction correction methods
such as those based on the Whitford reddening curve, provide
sufficiently good extinction correction to allow the \NIIOII\ diagnostic
to be used for reliable abundance determinations.

\item As has been shown previously, the common abundance diagnostic
\R23\ depends strongly on ionization parameter, and the common
ionization parameter diagnostic \OIIIOII\ depends strongly on
abundance.  An iterative approach is often required to resolve these
dependancies.  Unlike previous methods, we provide techniques for
explicitly determining the ionization parameter, rather than including
this as a `correction factor' to the abundance diagnostic.

\item Due to the local maximum in \R23, for objects with abundances
between $8.4 <$~\OH~$< 8.8$ (ie $0.6 <$~log(\R23)~$< 1.0$), a
different diagnostic to ours in this range should be used to obtain a
more reliable abundance estimate.  For \OH $\ge 8.9$, our \R23\ method
delivers a slightly higher abundance than the comparison abundance, but
this shift is equal to if not less than that seen for the other two
\R23\ -based diagnostics (M91 and Z94) in this region.
For \OH$ \le 8.5$, our \R23\ method is much more reliable than the
other two \R23\ based methods. The C01 \NIISII\ method
is also reliable in this region, with a similar degree of scatter.

\item We have presented a new combined diagnostic based only on
three \R23\ diagnostics: ours, M91 and Z94.  This diagnostic
eliminates the systematic shift inherent to all three techniques used
on their own and significantly reduces the scatter or rms error.

\item The ionization parameter diagnostic [\ion{S}{3}]/[\ion{S}{2}]
is independant of abundance, enabling a non-iterative approach
to be used if \SIII\ and \SII\ are available in the spectrum.
However, our current models do not allow for reliable ionization
parameter diagnostics using [\ion{S}{3}]/[\ion{S}{2}], or abundance
determinations derived from \dS23.  We believe this is a result of
either the use of an incorrect sulfur to oxygen abundance ratio,
errors in the sulfur depletion factor used, or errors in fundamental
atomic data.  If the sulfur lines are the only strong lines available,
we recommend an empirical method such as that given in \citet{Diaz00}
or the modified Diaz \& P\'{e}rez-Montero diagnostic presented here
which significantly reduces the errors inherent in the Diaz \&
P\'{e}rez-Montero diagnostic alone.

\item Both the \NIISII\ and \NIIHa\ ratios strongly depend on
ionization parameter, so an ionization parameter diagnostic should be
used to aid abundance determinations when using these ratios.  Neither
diagnostic is very sensitive to abundance and should be used with
caution. In particular, \NIISII\ based diagnostics have a larger
degree of uncertainty than schemes based on other ratios such as
\NIIOII\ or \R23.

\item Finally, for spectra of \HII\ regions or star-forming galaxies
in which the \NII, \OII, \OIII, \SII\, and the Balmer lines of Hydrogen
 are available, we present a method using a combination of
techniques for optimally determining the abundance from these strong
lines alone.  This method takes advantage of the reliability of our
\NIIOII\ for the intermediate to high metallicity range, and our
\R23\ diagnostic combined with the \citet{Zaritsky94},
\citet{McGaugh91} and \citet{Charlot01} diagnostics for the lower
abundance range.  This technique can be used over the entire
metallicity range, and appears to be prone to much smaller intrinsic
errors than all other techniques presented, including those of
\citet{McGaugh91,Zaritsky94,Charlot01}.  In addition, there is no
systematic offset in the derived abundance when compared with the average
of three previous techniques.  We strongly recommend use of this
technique if all the required emission-lines are available, and fluxes
can be extinction corrected using classical methods.

\end{enumerate}

\section{Acknowledgements}

This work has greatly benefited from discussions with 
Tim Heckman, Christy Tremonti,Margaret Geller, 
Rob Kennicutt,  Michael Strauss, and Lei Hao.
We thank Claus Leitherer and Brigitte Rocca-Volmerange for the
use of STARBURST99 and PEGASE 2, and we thank the referee for his
comments which made this a more thorough paper.  
L. Kewley gratefully acknowledges
support from the Australian Academy of Science Young Researcher
Scheme and from the French Service Culturel \& Scientifique.
M. Dopita acknowledges the support of the Australian National
University and the Australian Research Council through his
ARC Australian Federation Fellowship, and under the ARC Discovery
project DP0208445.

\clearpage

\clearpage

\clearpage

\clearpage

\clearpage

\clearpage

\clearpage

\clearpage

\clearpage

\clearpage

\clearpage

\clearpage

\clearpage

\clearpage

\clearpage

\clearpage

\clearpage

\clearpage

\clearpage

\clearpage

\clearpage

\clearpage

\newpage
\begin{deluxetable}{lrr}
\tabletypesize{\small}
\tablecaption{Solar metallicity ($Z_{\odot}$) and depletion factors (D) adopted
for each element.\label{Z_table}}
\tablehead{
\colhead{Element}
& \colhead{$\log({\rm Z_{\odot}})$}
& \colhead{$\log({\rm D})$}\\
}
\startdata
H & 0.00 & 0.00 \\
He & -1.01 & 0.00 \\
C & -3.44 & -0.30 \\
N & -3.95 & -0.22 \\
O & -3.07 & -0.22 \\
Ne & -3.91 & 0.00 \\
Mg & -4.42 & -0.70 \\
Si & -4.45 & -1.00 \\
S & -4.79 & 0.00 \\
Ar & -5.44 & 0.00 \\
Ca & -5.64 & -2.52 \\
Fe & -4.33 & -2.00 \\
\enddata
\end{deluxetable}
\begin{deluxetable}{llrrrrrrrr}
\tabletypesize{\scriptsize}
\tablecaption{Coefficients for Ionization Parameter Diagnostics
\label{qcoefftable}}
\tablehead{{Diagnostic (R)}
&
& {Z=0.05}
& {0.1}
& {0.2}
& {0.5}
& {1.0}
& {1.5}
& {2.0}
& {3.0 \Zsun}\\}
\startdata
$\log(\frac{[{\rm OIII}]}{[{\rm OII}]})$ & $k_{0}$ &  -36.9772 &  -74.2814
       & -36.7948 & -81.1880 &  -52.6367 & -86.8674 &  -24.4044 &  49.4728 \\
   &  $k_{1}$ &  10.2838  & 24.6206 &  10.0581 & 27.5082 &  16.0880 & 28.0455
       &   2.51913 & -27.4711 \\
   &  $k_{2}$ &  -0.957421 & -2.79194 & -0.914212 & -3.19126 & -1.67443
       & -3.01747 &  0.452486 &  4.50304 \\
   &  $k_{3}$ & 0.0328614 & 0.110773 & 0.0300472 & 0.128252 & 0.0608004
       & 0.108311 & -0.0491711 & -0.232228 \\ \hline
$\log(\frac{[{\rm OIII}]}{[{\rm OII}]})$ & \multicolumn{7}{l}{
(Combined method, Section~\ref{Combined_method})}\\
  & $k_{0}$ & 7.39167 & 7.46218 & 7.57817 & 7.73013 &  \nodata & 
\nodata &  \nodata &  \nodata \\
   &  $k_{1}$ & 0.667891  & 0.685835 & 0.739315 & 0.843125 & \nodata & 
\nodata  & \nodata  & \nodata \\
   &  $k_{2}$ & 0.0680367 & 0.0866086 & 0.0843640 & 0.118166 & \nodata 
& \nodata & \nodata & \nodata \\ \hline
$\log(\frac{[{\rm SIII}]}{[{\rm SII}]})$ & $k_{0}$ &  30.0116 &   16.8569
       &  32.2358 &  -3.06247  & -2.94394 & -38.1338  &-21.0240  & -6.61131 \\
   &  $k_{1}$ & -14.8970 &  -9.62876 & -15.2438 & -0.864092 & -0.546041
       &  13.0914 &  6.55748  &  1.36836  \\
   &  $k_{2}$ &    2.30577 &   1.59938 &  2.27251 &   0.328467 &  0.239226
       &  -1.51014 & -0.683584 & -0.0717560\\
   &  $k_{3}$ & -0.112314 & -0.0804552 & -0.106913 & -0.0196089 & -0.0136716
       & 0.0605926 & 0.0258690 & 0.00225792\\ \hline
\enddata
\end{deluxetable}

\begin{deluxetable}{llrrrrrrr}
\tabletypesize{\scriptsize}
\tablecaption{Coefficients for Abundance Diagnostics
\label{coefftable}}
\tablehead{{Diagnostic (R)}
&
& {$q=5\times10^{6}$}
& {$1\times10^{7}$}
& {$2\times10^{7}$}
& {$4\times10^{7}$}
& {$8\times10^{7}$}
& {$1.5\times10^{8}$}
& {$3.0\times10^{8}$} \\}
\startdata
$\log(\frac{[{\rm NII}]}{[{\rm OII}]})$ & $k_{0}$ & 616.294 & 859.253
     & 1106.87 & 1307.93 & 1270.42 & 1067.21 & 751.533 \\
& $k_{1}$ &  -298.819 &-413.604 &-532.154 &-628.828 &-612.566 &-517.101
     &-367.682 \\
& $k_{2}$ &  54.7919 & 75.1020 & 96.3733 & 113.802 & 111.214 & 94.4377
     & 67.9579 \\
& $k_{3}$ & -4.51877 &-6.11475 &-7.81061 &-9.20734 &-9.02939 &-7.72256
     & -5.64034 \\
& $k_{4}$ & 0.141576 &0.188586 &0.239282 &0.281264 &0.276854 &0.238784
     &0.177489 \\ \hline
$\log(\frac{[{\rm NII}]}{[{\rm SII}]})$ & $k_{0}$ &-1042.47 & -1879.46
     & -2027.82 & -2080.31 & -2162.93 & -2368.56 & -2910.63\\
& $k_{1}$ &   521.076 &  918.362 &  988.218  & 1012.26  & 1048.97
     &  1141.97  & 1392.18 \\
& $k_{2}$ &   -97.1578 & -167.764 & -180.097 & -184.215 & -190.260
     &  -205.908 & -249.012 \\
& $k_{3}$ &   8.00058 &   13.5700  &  14.5377 &   14.8502 &   15.2859
     &   16.4451 &   19.7280 \\
& $k_{4}$ & -0.245356 &-0.409872 &-0.438345 &-0.447182 &-0.458717
     &-0.490553 &-0.583763 \\ \hline
$\log$(\R23) &  $k_{0}$ &  -3267.93 & -3727.42 & -4282.30 & -4745.18
& -4516.46 & -3509.63 & -1550.53 \\
&  $k_{1}$ &   1611.04 &  1827.45 &  2090.55 &  2309.42 &  2199.09
     &  1718.64 &  784.26 \\
&  $k_{2}$ &  -298.187 & -336.340 & -383.039  &-421.778 & -401.868
     & -316.057 & -149.245 \\
&  $k_{3}$ &  24.5508  & 27.5367  &31.2159 &  34.2598  & 32.6686
      & 25.8717  & 12.6618 \\
&  $k_{4}$ & -0.758310 &-0.845876 &-0.954473 & -1.04411 &-0.996645
      & -0.795242 &-0.403774 \\ \hline
$\log$(${\rm S}_{23}$) &  $k_{0}$ &  -1543.68 & -1542.15 & -1749.48 & -1880.06
      & -1627.10 & -1011.65&   -81.6519 \\
&  $k_{1}$ &   761.018 &  758.664 &  855.280 &  914.362 &   790.891
      &  497.017  &   55.3453 \\
&  $k_{2}$ &  -141.061 & -140.351 & -157.198 & -167.192 &  -144.699
      & -92.2429 &  -13.7783 \\
&  $k_{3}$ &   11.6389 &  11.5597 & 12.8626 & 13.6119  &  11.7988
      & 7.64915 &  1.46716 \\
&  $k_{4}$ & -0.360280 & -0.357261 & -0.394978 & -0.415997  &-0.361423
      &-0.238660 &-0.0563760\\ \hline
$\log(\frac{[{\rm NII}]}{{\rm H}\alpha})$ & $k_{0}$ &-2700.08 &-2777.11
      &-2940.90 & -3073.05 &-3049.60 &-2983.69 & -3100.57 \\
&  $k_{1}$ &  1335.14 & 1369.97 &1445.50 &  1505.94&  1491.07 &  1454.45
      & 1501.77 \\
&  $k_{2}$ &  -247.533 &-253.434 &-266.482 & -276.829 &-273.510 & -266.015
      &  -272.883 \\
&  $k_{3}$ &   20.3663&  20.8100 & 21.8103 & 22.5946 &  22.2767 & 21.6024 
      &  22.0132 \\
&  $k_{4}$ & -0.62692 & -0.63942 & -0.6681 & -0.6903 & -0.6792 & -0.6566
      &  -0.6646 \\ \hline
$\log(\frac{[{\rm NII}]}{[{\rm OIII}]})$  & $k_{0}$ &   912.833
      &     3720.98  &    4180.19    &  4289.18    &  4209.16    &  4013.22
      &     3246.13 \\
   & $k_{1}$ &     -461.733   &  -1792.96   &  -2011.36   &  -2064.05
      &   -2032.26  &   -1950.99   &  -1604.38 \\
   & $k_{2}$ &      87.8445   &   324.052   &   362.929  &    372.477
      & 367.999   &    355.774    &  297.497 \\
   & $k_{3}$ &  -7.45740  &   -26.0543  &  -29.1288 & -29.9016
      & -29.6506    & -28.8754 &   -24.5628 \\
   & $k_{4}$ & 0.238581   & 0.786768  &  0.877939  &  0.901534
      &   0.897467  & 0.880623 &  0.762337 \\ \hline
$\log(\frac{[{\rm NII}]}{[{\rm OII}]})$ & \multicolumn{7}{l}{
(Combined method, Section~\ref{Combined_method})} \\
& $k_{0}$ & 1.49089 & 1.51444 & 1.54020 & 1.55753 & 1.56152 & 1.55795 
& 1.55354 \\
& $k_{1}$ & 1.34637 & 1.31116 & 1.26602 & 1.22723 & 1.20101 & 1.19010 
& 1.19062 \\
& $k_{2}$ & 0.204543 & 0.191521 & 0.167977 & 0.145826 & 0.132789 & 
0.131257  & 0.139017 \\ \hline
$\log$(\R23) & \multicolumn{7}{l}{
(Combined method, Section~\ref{Combined_method})} \\
& $k_{0}$ & -27.0004 & -31.2133 & -36.0239 & -40.9994 & -44.7026 & 
-46.1589 &-45.6075  \\
& $k_{1}$ & 6.03910 &7.15810  & 8.44804 & 9.78396 & 10.8052 & 11.2557 
& 11.2074 \\
& $k_{2}$ & -0.327006 & -0.399343 & -0.483762 & -0.571551  & 
-0.640113 &  -0.672731 & -0.674460 \\  \hline
\enddata
\end{deluxetable}

\begin{deluxetable}{lllrl}
\tabletypesize{\scriptsize}
\tablecaption{RMS Error measured from the solid $x=y$ line shown on 
the Figures given in column 1.
\label{rms_table}}
\tablehead{
Figure & X-axis & Y-axis & RMS Error & Comments \\}
\startdata
\ref{M91_vs_Z94_vs_C01}a &  M91 & Z94 & 0.18 & strong systematic shift,
tight correlation at high abundances \\
\ref{M91_vs_Z94_vs_C01}b & M91 & C01 & 0.11 & large degree of scatter,
systematic shift\\
\ref{M91_vs_Z94_vs_C01}c & C01 & Z94 & 0.13 & large degree of scatter,
small systematic shift\\
\ref{M91_Z94_C01_vs_ave}a & Comp. Ave & M91 & $>0.09$ & M91 
contributes to both axes, systematic shift.  \\
\ref{M91_Z94_C01_vs_ave}b & Comp. Ave & Z94 & $>0.10$ & M91 
contributes to both axes, systematic shift.  \\
\ref{M91_Z94_C01_vs_ave}c & Comp. Ave & C01 & $>0.06$ & C01 
contributes strongly to both axes \\
\ref{M91Z94ave_vs_C01} & M91,Z94 Ave & C01 & 0.09 & no systematic shift  \\
\ref{KNIIOII_ave} & Comp. Ave & KD - [NII]/[OII] & 0.04 & no systematic shift.\\
\ref{KNIIOII_ave_extinct} & Comp. Ave & KD - [NII]/[OII] reddened & 0.14 
&systematic shift, large degree of scatter\\ 
\ref{KR23_ave} &  Comp. Ave & KD - \R23 & 0.14 & systematic shift, 
large scatter at low abundances.\\
\ref{KNIISII_ave} & Comp. Ave & KD - [NII]/[SII] & 0.18 & strong 
systematic shift and scatter. \\
\ref{qOIIIOII_vs_qSIIISII} & KD - [SIII]/[SII] & KD - [OIII]/[OII] & 0.25
& strong scatter \& systematic shift \\
\ref{S23_vs_compave} &  Comp. Ave & KD - \dS23 & 0.32 & strong 
systematic shift and scatter \\
\ref{OtherAve_vs_DiazS23} & Comp. Ave & DP00 \dS23 & 0.44 & large scatter, systematic
shift \\
\ref{OtherAve_vs_DiazS23_2} & Comp. Ave & DP00/KD \dS23 & 0.25 & smaller scatter
than DP00, but still significant \\
\ref{Combined_vs_otherave} & Comp. Ave & KD - combination scheme & 
0.05 & no systematic shift, 4 outliers
(rms is 0.04 without outliers)\\
\ref{OtherAve_vs_R23only} &  Comp. Ave & \R23 only combined scheme & 
$>0.06$ & no systematic shift \\
\enddata
\end{deluxetable}

\end{document}